\documentclass[12pt]{article}
\usepackage{amssymb}
\usepackage{amsmath}
\usepackage{IEEEtrantools}
\usepackage{pdflscape}
\usepackage{afterpage}
\usepackage{capt-of}

\usepackage{graphicx}
\usepackage{epstopdf}
\usepackage{subfig}

\numberwithin{equation}{section}

\newcommand{\be}{\begin{eqnarray}}
\newcommand{\ee}{\end{eqnarray}}
\newcommand{\non}{\nonumber}
\newcommand{\id}{\mathbb{I}}
\newcommand{\tr}{\mathop{\rm tr}\nolimits}
\newcommand{\diag}{\mathop{\rm diag}\nolimits}
\newcommand{\uh}{\frac{u}{2}}
\newcommand{\half}{\frac{1}{2}}

\begin{document}

\begin{titlepage}
\strut\hfill UMTG--290
\vspace{.5in}
\begin{center}

\LARGE Quantum group symmetries and completeness\\
for $A_{2n}^{(2)}$ open spin chains\\
\vspace{1in}
\large Ibrahim Ahmed \footnote{ibrahimahmed@miami.edu}, Rafael I. Nepomechie \footnote{nepomechie@miami.edu}
and Chunguang Wang \footnote{c.wang22@umiami.edu}\\[0.8in]
\large Physics Department, P.O. Box 248046, University of Miami\\[0.2in]  
\large Coral Gables, FL 33124 USA\\

\end{center}

\vspace{.5in}

\begin{abstract}
We argue that the Hamiltonians for $A^{(2)}_{2n}$ open quantum spin
chains corresponding to two choices of integrable boundary conditions
have the symmetries $U_{q}(B_{n})$ and $U_{q}(C_{n})$, respectively.
We find a formula for the Dynkin labels of the Bethe states (which
determine the degeneracies of the corresponding eigenvalues) in terms
of the numbers of Bethe roots of each type.  With the help of this
formula, we verify numerically (for a generic value of the anisotropy
parameter) that the degeneracies and multiplicities of the spectra
implied by the quantum group symmetries are completely described by
the Bethe ansatz.
\end{abstract}

\end{titlepage}

\setcounter{footnote}{0}

\section{Introduction}\label{sec:intro}

Interesting new connections of integrable quantum spin chains 
to integrable quantum field theory, conformal field 
theory (CFT) and string theory, as well as to condensed matter 
physics, continue to be found. A case in point 
concerns the $A^{(2)}_{n}$ family of models \cite{Izergin:1980pe, 
Bazhanov:1984gu, Bazhanov:1986mu, Jimbo:1985ua}, which has recently 
been revisited by Vernier {\em et al.} \cite{Vernier:2014uja, Vernier:2014b, Vernier:2016xha}. 
For example, it was argued in \cite{Vernier:2014uja} 
that the $A^{(2)}_{2}$ model \cite{Izergin:1980pe} has a regime where the continuum limit 
is a certain non-compact CFT, the so-called black hole sigma model 
\cite{Witten:1991yr, Dijkgraaf:1991ba}.

Another interesting feature of these models is that they can have
quantum group symmetries (see e.g. \cite{Kulish:1983md,
Chari:1994pz}), provided that the boundary conditions are suitable.
For the closed chains with periodic boundary conditions studied in
\cite{Vernier:2014uja, Vernier:2014b, Vernier:2016xha}, such
symmetries can be realized only indirectly; however, quantum group
symmetries can be realized directly in open chains
\cite{Pasquier:1989kd}.

Motivated in part by these recent developments,
we have set out to revisit the quantum group symmetries of the $A^{(2)}_{n}$
family of models.  We therefore focus instead on {\em open}
chains; and, for concreteness, we restrict here to the even series
$A^{(2)}_{2n}$, leaving the odd series $A^{(2)}_{2n-1}$ for a future publication.
It has long been known that, for one simple set of integrable
boundary conditions, the former models have $U_{q}(B_{n})$ symmetry 
\cite{Mezincescu:1990ui, Mezincescu:1991rb}. 

We argue here that -- surprisingly -- the $A^{(2)}_{2n}$ models have $U_{q}(C_{n})$ symmetry
for another set of integrable boundary conditions. (The symmetry 
for the case $n=1$ was already noticed in \cite{Nepomechie:1999jz}, but the 
symmetry for the general case $n>1$ had remained unexplored until 
now.) 
The symmetries (both $U_{q}(B_{n})$ and $U_{q}(C_{n})$)
determine the degeneracies and multiplicities of the spectra, which
are completely described by the Bethe ansatz solutions.

The outline of this paper is as follows. In Section \ref{sec:models} 
we briefly review the construction of the integrable $A_{2n}^{(2)}$ open quantum spin
chains that are the focus of this paper. In Section \ref{sec:simple} 
we show that the Hamiltonians for the two cases of interest
can be expressed as sums of two-body 
terms. We use this fact in Section \ref{sec:qg}
to demonstrate that the Hamiltonians have quantum group symmetries, 
which in turn determine the degeneracies and 
multiplicities of the spectra. In Section \ref{sec:BA} we briefly review the Bethe 
ansatz solutions of the models, and we obtain a formula for the Dynkin labels of the
Bethe states, part of whose proof is sketched in an appendix. In 
Section \ref{sec:completeness} we use this formula to help verify 
numerically that the Bethe ansatz solutions completely account for 
the degeneracies and multiplicities implied by the quantum group symmetries.
In Section \ref{sec:conclusion} we briefly summarize our conclusions, 
and list some interesting open problems.

\section{The models}\label{sec:models}

We briefly review here the construction of the integrable $A_{2n}^{(2)}$ open quantum spin
chains that will turn out to have quantum group symmetries. The basic ingredients are the 
R-matrix and K-matrices, which are used to construct a commuting transfer 
matrix that contains the integrable Hamiltonian.

\subsection{R-matrix}

The R-matrix is a matrix-valued function $R(u)$ of the so-called spectral parameter 
$u$ that maps ${\cal V} \otimes {\cal V}$ to itself, where here
${\cal V}$ is a $(2n+1)$-dimensional vector space, which is a solution of the 
Yang-Baxter equation (YBE) on ${\cal V} \otimes {\cal V}\otimes {\cal V}$
\be
R_{12}(u - v)\,  R_{13}(u)\, R_{23}(v) = R_{23}(v)\,  R_{13}(u)\, R_{12}(u - v)
\,.  \label{YBE}
\ee
We use the standard notations $R_{12} = R \otimes \id\,, R_{23} = \id \otimes R\,, R_{13} = 
{\cal P}_{23}  R_{12} {\cal P}_{23}$, where $\id$ is the identity 
matrix on ${\cal V}$, and ${\cal P}$ is the permutation matrix on ${\cal V} \otimes {\cal V}$
\be
{\cal P}=\sum_{\alpha, \beta = 1}^{2n+1} e_{\alpha \beta}\otimes e_{\beta 
\alpha} \,,
\ee
and $e_{\alpha \beta}$ are the $(2n+1) \times (2n+1)$ elementary 
matrices with elements $(e_{\alpha \beta})_{ij} = 
\delta_{\alpha, i} \delta_{\beta, j}$. 

We focus here on the R-matrix (\ref{Rmatrix})
that is associated with the fundamental representation of
$A_{2n}^{(2)}$ \cite{Bazhanov:1984gu, Bazhanov:1986mu, Jimbo:1985ua} 
with anisotropy parameter $\eta$, 
which is a generalization of the Izergin-Korepin R-matrix 
\cite{Izergin:1980pe} that is associated with $A_{2}^{(2)}$. Besides 
satisfying the YBE, this R-matrix enjoys several additional important properties, among 
them $PT$ symmetry
\be
R_{21}(u) \equiv {\cal P}_{12}\, R_{12}(u)\, {\cal P}_{12} 
= R_{12}^{t_1 t_2}(u) \,,
\label{PT}
\ee
unitarity
\be
R_{12}(u)\ R_{21}(-u) = \xi(u)\, \xi(-u)\, \id\otimes\id  \,,
\label{unitarity}
\ee
where $\xi(u)$ is given by
\be
\xi(u)=2\sinh(\uh -2\eta) \cosh(\uh -(2n+1)\eta) \,,
\label{xi}	    
\ee
regularity
\be
R(0) = \xi(0)\, {\cal P}\,,
\label{Rregular}
\ee
and crossing symmetry
\be
R_{12}(u)=V_1\, R_{12}^{t_2}(-u-\rho)\, V_1
= V_2^{t_2}\, R_{12}^{t_1}(-u-\rho)\, V_2^{t_2} \,,
\label{crossing}
\ee
where $\rho=-i \pi -2(2n+1)\eta$; and the matrix $V$, which is given by 
(\ref{Vmat}), satisfies $V^2 = \id$.

\subsection{K-matrices}

The matrix $K^{-}(u)$, which maps ${\cal V}$ to itself, is a solution of the 
boundary Yang-Baxter equation (BYBE) on ${\cal V} \otimes {\cal V}$ 
\cite{Cherednik:1985vs, Sklyanin:1988yz, Mezincescu:1990uf, Ghoshal:1993tm}
\be
R_{12}(u - v)\, K^-_1(u)\ R_{21} (u + v)\, K^-_2(v)
= K^-_2(v)\, R_{12}(u + v)\, K^-_1(u)\, R_{21}(u - v)  \,.
\label{BYBEm}
\ee
The matrix $K^{-}(u)$ is assumed to have the regularity property
\be
K^{-}(0) = \kappa\, \id \,.
\label{Kregular}
\ee
Similarly, $K^{+}(u)$ satisfies \cite{Sklyanin:1988yz, Mezincescu:1990uf}
\be
\lefteqn{R_{12}(-u + v)\, K_1^{+\, t_1}(u)\, M^{-1}_1\, R_{21} (-u -v 
-2\rho)\, M_1\, K_2^{+\, t_2}(v)} \non \\
& & \quad = K^{+\, t_2}_2(v)\, M_1\, R_{12}(-u - v- 2\rho)\, M^{-1}_1\,
K^{+\, t_1}_1(u)\, R_{21}(-u +v)  \,, 
\label{BYBEp}
\ee 
where the matrix $M$ is defined by
\be
M = V^{t}\, V\,,
\label{Mdef}
\ee
and is given by (\ref{Mmat}). If $K^{-}(u)$ is a solution of the BYBE
(\ref{BYBEm}), then \cite{Sklyanin:1988yz, Mezincescu:1990uf}
\be
K^{+}(u) = K^{-\, t}(-u-\rho)\, M
\label{isomorphism}
\ee
is a solution of (\ref{BYBEp}). 

We consider here two different sets of K-matrices:
\be
&(I):& \qquad K^{-}(u) = \id\,, \qquad\quad K^{+}(u) = M \,, 
\label{KsetI} \\
&(II):& \qquad K^{-}(u) = K(u)\,, \quad K^{+}(u) = K(-u-\rho)\, M \,.
\label{KsetII}
\ee
The fact that $K^{-}(u) = \id$ is a solution of the BYBE was noted in 
\cite{Mezincescu:1990ui}. The matrix $K(u)$ in (\ref{KsetII}) is 
the diagonal matrix given by
\be
K(u) = \diag( k_{1}(u), \ldots, k_{2n+1}(u) ) \,,
\label{Kmat}
\ee
where
\be
k_{j}(u) = \left\{ \begin{array}{ll}
e^{-u} \left[ \epsilon i \cosh \eta + \sinh(u- 2 n \eta) \right] & j = 1, 
\ldots, n \\
\epsilon i \cosh(u + \eta) - \sinh(2 n \eta) & j = n+1 \\
e^{u} \left[ \epsilon i\cosh \eta + \sinh(u- 2 n \eta) \right] & j = n+2, 
\ldots, 2n+1
\end{array} \right. \,,
\label{Kmat2}
\ee 
where $\epsilon$ can have the values $\pm 1$, but for concreteness we 
henceforth set $\epsilon = +1$. This K-matrix has the regularity property (\ref{Kregular}) with
\be
\kappa = i \cosh \eta - \sinh(2 n \eta) \,.
\label{kappa}
\ee
The solution (\ref{Kmat})-(\ref{Kmat2}) of the BYBE  (\ref{BYBEm}) for the case $n=1$ was found in 
\cite{Mezincescu:1990ui}, and the generalization for $n>1$ was 
found in \cite{Batchelor:1996np, LimaSantos:2002ui}.

\subsection{Transfer matrix and Hamiltonian}

The transfer matrix $t(u)$ for an integrable open quantum spin chain with $N$ 
sites, which acts on the quantum space ${\cal V}^{\otimes N}$, is given by \cite{Sklyanin:1988yz}
\be
t(u) = \tr_a K^{+}_a(u)\, T_a(u)\,  K^{-}_{a}(u)\, \hat T_a(u) \,, 
\label{transfer}
\ee
where the monodromy matrices are defined by
\be 
T_a(u) = R_{aN}(u)\ R_{a N-1}(u)\ \cdots R_{a1}(u) \,,  \qquad
\hat T_a(u) = R_{1a}(u)\ \cdots R_{N-1 a}(u)\ R_{Na}(u) \,,  
\label{monodromy}
\ee
and the trace in (\ref{transfer}) is over the auxiliary space, which 
we denote by $a$. The various properties satisfied by the R and K 
matrices can be used to show that the transfer matrix satisfies the 
fundamental commutativity property \cite{Sklyanin:1988yz}
\be
\left[ t(u) \,, t(v) \right] = 0 \hbox{   for all   } u \,, v \,.
\label{commutativity}
\ee

The corresponding integrable open chain Hamiltonian ${\cal H}$ is given (up to 
multiplicative and additive constants) by $t'(0)$, which evidently 
satisfies
\be
\left[ {\cal H} \,, t(u) \right] = 0 \,.
\ee
More explicitly, one finds \cite{Sklyanin:1988yz}
\be
{\cal H} = \sum_{k=1}^{N-1} h_{k,k+1} +\frac{1}{2\kappa} K^{-\, 
'}_{1}(0) + 
\frac{1}{\tr K^{+}(0)}\tr_{0} K^{+}_{0}(0) h_{N0}\,,
\label{Hamiltonian}
\ee
where the two-site Hamiltonian $h_{k,k+1}$ is given by
\be
h_{k,k+1} = \frac{1}{\xi(0)} {\cal P}_{k, k+1} R'_{k,k+1}(0)  \,.
\label{twosite}
\ee

\section{Simplification of the Hamiltonian}\label{sec:simple}

We show here that the boundary terms in the Hamiltonian 
(\ref{Hamiltonian}) can be simplified for the two sets of K-matrices 
(\ref{KsetI}), (\ref{KsetII}) in such a way that the Hamiltonians are 
expressed as sums of two-body terms, which will allow us to demonstrate their quantum 
group invariance in the following section. The 
key step in this simplification is a K-matrix identity (\ref{identity}), which is 
reminiscent of Sklyanin's ``less obvious'' isomorphism given by Eqs. (17)
and (18) in \cite{Sklyanin:1988yz}, and the Ghoshal-Zamolodchikov
boundary crossing-unitarity relation, see
Eqs. (3.33) and (3.35) in  \cite{Ghoshal:1993tm}.

\subsection{An identity for the K-matrix}

A useful identity is
\be
\tr_{1} K^{-}_{1}(-u-\rho)\, M_{1}\, R_{12}(2u)\, {\cal 
P}_{12}  = f(u)\, V_{2}\, K^{-\, t_{2}}_{2}(u)\, V_{2} \,,
\label{identity}
\ee
where $f(u)$ is a scalar function. The remainder of this subsection is devoted to proving this identity.
Readers who are more interested to see how this identity can be used
to simplify the boundary terms in the Hamiltonian may skip directly to
Sec.  \ref{sec:ham}.

It is helpful to recall (see e.g. 
\cite{Mezincescu:1991ke}) that the crossing symmetry (\ref{crossing})
can be used to show that the R-matrix degenerates at $u=-\rho$ to a 
projector onto a one-dimensional subspace,
\be
\tilde P^{-}_{12} \equiv \frac{1}{(2n+1)\, \xi(0)} R_{12}(-\rho) = 
\frac{1}{(2n+1)} V_{1}\, {\cal P}^{t_{2}}_{12}\, V_{1} \,,
\label{Ptilde}
\ee
which obeys
\be
\left( \tilde P^{-}_{12} \right)^{2} = \tilde P^{-}_{12} 
\label{proj1}
\ee
and 
\be
\tilde P^{-}_{12}\, A_{12}\, \tilde P^{-}_{12} = \tr_{12} \left( 
\tilde P^{-}_{12} A_{12}\right) \tilde P^{-}_{12} \,,
\label{proj2}
\ee
where $A$ is an arbitrary matrix acting on ${\cal V}\otimes {\cal 
V}$. This projector is not symmetric,
\be
\tilde P^{-}_{21} \equiv {\cal P}_{12}\, \tilde P^{-}_{12}\, {\cal P}_{12} 
= (\tilde P^{-}_{12})^{t_{1} t_{2}} \ne \tilde P^{-}_{12} \,.
\ee
We also recall that 
\be
V_{1}\, R_{12}(u)\, V_{1} = V_{2}\, R_{21}(u)\, V_{2} \,.
\label{useful}
\ee

The starting point of the proof is the BYBE (\ref{BYBEm}), where we set 
$v=-u-\rho$ and use the definition (\ref{Ptilde}) to obtain
\be
R_{12}(2u+\rho)\, K^-_1(u)\, \tilde P^{-}_{21}\, K^-_2(-u-\rho)
= K^-_2(-u-\rho)\, \tilde P^{-}_{12}\, K^-_1(u)\, R_{21}(2u +\rho)  \,.
\ee
With the help of the relations 
\be
\tilde P^{-}_{21} = V_{1}^{t_{1}}\, V_{2}^{t_{2}}\, \tilde P^{-}_{12}\,  V_{1}^{t_{1}}\, V_{2}^{t_{2}}
\ee
and
\be
R_{21}(2u +\rho) = V_{1}^{t_{1}}\, V_{2}^{t_{2}}\, R_{12}(2u +\rho)\, 
V_{1}^{t_{1}}\, V_{2}^{t_{2}}
\ee
that follow from (\ref{useful}), we arrive at
\be
\lefteqn{R_{12}(2u+\rho)\, K^-_1(u)\,  V_{1}^{t_{1}}\, V_{2}^{t_{2}}\, \tilde P^{-}_{12}\, V_{2}^{t_{2}}
\, K^-_2(-u-\rho)}\non\\
&&= K^-_2(-u-\rho)\, \tilde P^{-}_{12}\, K^-_1(u)\, 
V_{1}^{t_{1}}\, V_{2}^{t_{2}}\, R_{12}(2u +\rho)\, V_{2}^{t_{2}} \,.
\ee
Multiplying both sides on the right by $\tilde P^{-}_{12}$ and using 
the projector property (\ref{proj2}), we obtain
\be
R_{12}(2u+\rho)\, K^-_1(u)\,  V_{1}^{t_{1}}\, V_{2}^{t_{2}}\, \tilde 
P^{-}_{12} = g(u)\, K^-_2(-u-\rho)\, \tilde P^{-}_{12} \,,
\ee
where $g(u)$ is some scalar function. Multiplying both sides, on both 
the right and the left, by the permutation matrix ${\cal P}_{12}$, 
and then using the crossing equation (\ref{crossing}) and the 
expression (\ref{Ptilde}) for $\tilde P^{-}_{12}$, we obtain
\be
V_{1}^{t_{1}}\, R_{12}^{t_{1}}(-2u-2\rho)\, K^-_2(u)\, 
V_{1}^{t_{1}}\, V_{2}^{t_{2}}\, {\cal P}^{t_{1}}_{12}\, V_{1}^{t_{1}}
= g(u)\, K^-_1(-u-\rho)\, V_{1}^{t_{1}}\, {\cal P}^{t_{1}}_{12}\, 
V_{1}^{t_{1}}\,.
\ee
Taking the trace of both sides over the 
first space, we arrive at
\be
\tr_{1} R_{12}^{t_{1}}(-2u-2\rho)\, K^-_2(u)\, 
V_{1}^{t_{1}}\, V_{2}^{t_{2}}\, {\cal P}^{t_{1}}_{12}
= g(u)\, \tr_{1} K^-_1(-u-\rho)\,  V_{1}^{t_{1}}\, {\cal P}^{t_{1}}_{12}\, V_{1}^{t_{1}}\,,
\ee
which can be simplified to
\be
\tr_{1}  K^-_1(u)\, M_{1}\, 
R_{12}(-2u-2\rho)\, {\cal P}_{12}
= g(u)\, V_{2}\, K^{-\, t_{2}}_{2}(-u-\rho)\, V_{2}\,.
\ee
Replacing $u \mapsto -u-\rho$ and setting $f(u) = g(-u-\rho)$, 
we finally obtain (\ref{identity}).

\subsection{Simplified Hamiltonians}\label{sec:ham}

We now proceed to simplify the boundary terms in the Hamiltonian (\ref{Hamiltonian})
using the identity  (\ref{identity}), which can be rewritten as
\be
\tr_{1} K^{+}_{1}(u)\, {\cal P}_{12} R_{21}(2u)   = f(u)\, V_{2}\, K^{-}_{2}(u)\, V_{2} 
\label{identity2}
\ee
for diagonal $K^{\pm}$-matrices that are related by (\ref{isomorphism}).

\subsubsection{Set I}

For the first set of K-matrices (\ref{KsetI}), the identity 
(\ref{identity2}) immediately implies that
\be
\tr_{1} M_{1}\, {\cal P}_{12} R_{21}(2u)   = f(u)\, \id_{2}\,.
\label{identity3a}
\ee
Differentiating this relation with respect to $u$ and then setting 
$u=0$, we obtain the result
\be
\tr_{1} M_{1}\, {\cal P}_{12} R'_{21}(0) \propto \id_{2}
\label{identity3b}
\ee
(see also \cite{Mezincescu:1990ui, Reshetikhin:1990})
and therefore
\be
\tr_{0} K^{+}_{0}(0) h_{N0} = \tr_{0} M_{0} h_{N0} \propto \tr_{0} M_{0} 
{\cal P}_{N 0} R'_{N 0}(0) \propto \id_{N} \,,
\ee 
i.e. the corresponding boundary term is proportional to the identity 
matrix. Moreover, since 
$K^{-}(u)=\id$, the boundary term with $K^{-\, '}(0)$ evidently vanishes.

In short, the two boundary terms in the expression 
(\ref{Hamiltonian}) for the Hamiltonian can be dropped. 
The Hamiltonian for the set I therefore reduces to a sum of two-site Hamiltonians  \cite{Mezincescu:1990ui}
\be
{\cal H}^{(I)} = \sum_{k=1}^{N-1} h_{k,k+1} \,.
\label{HamiltonianI}
\ee
Its relation to the transfer matrix (\ref{transfer}) is given by
\be
{\cal H}^{(I)} = \frac{1}{c_{1}} t'(0) + c_{2} \id^{\otimes N}\,,
\label{HIt}
\ee
with
\be
c_{1} &=& 4^{N+1}  \sinh((2n+1)\eta)\, 
\cosh((2n-1)\eta)\, \sinh^{2N-1}(2\eta)\, \cosh^{2N}((2n+1)\eta)\,, \non\\
c_{2} &=&\frac{\cosh((6n+1)\eta)}{2 \sinh((4n+2)\eta)\, 
\cosh((2n-1)\eta)} \,.
\label{HItc1c2}
\ee
The Hamiltonian (\ref{HamiltonianI}) is Hermitian for real, but not 
for imaginary, values of $\eta$.

\subsubsection{Set II}

We turn now to the second set of K-matrices (\ref{KsetII}). 
Setting $u=0$ in the identity (\ref{identity2}), and using the 
regularity properties (\ref{Rregular}) and (\ref{Kregular}), 
we obtain
\be
f(0) = \frac{1}{\kappa}\, \xi(0) \tr K^{+}(0) \,.
\label{f0}
\ee
Moreover, differentiating the identity (\ref{identity2}) with respect to $u$ and then setting $u=0$, 
we obtain
\be
2 \tr_{1} K^{+}_{1}(0)\, {\cal P}_{12} R'_{21}(0) + \ldots = f(0)\, V_{2}\,
K^{-\, '}_{2}(0) V_{2}\, + \ldots \,,
\label{intermed}
\ee
where the ellipses represent terms that are proportional to the 
identity, which we drop. Using the explicit form of 
the K-matrix (\ref{Kmat})-(\ref{Kmat2}), we observe that
\be
V\,  K^{-\, '}(0)\,  
V = - K^{-\, '}(0)  + \mu U +  \nu \id \,,
\label{observe}
\ee
where 
\be
\mu =  2(i \sinh \eta - \cosh(2 n \eta))\,, \qquad
\nu = 2\cosh(2 n \eta)\,, \qquad 
U=e_{n+1,n+1} \,.
\ee 
Substituting (\ref{f0}) and (\ref{observe}) into 
(\ref{intermed}), we arrive at the identity
\be
\frac{1}{\xi(0)\, \tr K^{+}(0)} \tr_{1} K^{+}_{1}(0)\, {\cal P}_{12} R'_{21}(0)
 = -\frac{1}{2\kappa}K^{-\, '}_{2}(0)  + \frac{\mu}{2\kappa} U_{2} + \ldots
\ee
The Hamiltonian (\ref{Hamiltonian}) for the set II therefore reduces to the form
\be
{\cal H}^{(II)} = \sum_{k=1}^{N-1} h_{k,k+1}
+ \frac{1}{2\kappa} \left[ K^{-\, '}_{1}(0) - K^{-\, '}_{N}(0) \right]
+ \frac{\mu}{2\kappa} U_{N} \,.
\ee
Let us define a new two-site Hamiltonian $\tilde h_{k,k+1}$ as follows
\be
\tilde h_{k,k+1} \equiv h_{k,k+1} + \frac{1}{2\kappa} \left[ K'_{k}(0)
-  K'_{k+1}(0) \right] \,.
\label{twositeII}
\ee
We conclude that, up to a term proportional to $U_{N}$,
the Hamiltonian again reduces to a sum of two-site 
Hamiltonians,
\be
{\cal H}^{(II)} = \sum_{k=1}^{N-1} \tilde h_{k,k+1} + 
\frac{\mu}{2\kappa} U_{N} \,.
\label{HamiltonianII}
\ee
Its relation to the transfer matrix (\ref{transfer}) is given by
\be
{\cal H}^{(II)} = \frac{1}{c_{1}} t'(0) + c_{2} \id^{\otimes N}\,,
\label{HIIt}
\ee
with
\be
c_{1} &=& 2^{2N+1}  (\cosh \eta+ i \sinh(2 n \eta))^{2}\, \sinh((4n+2)\eta)\, 
\cosh((2n+3)\eta)\, \left[\sinh(2\eta)\, 
\cosh((2n+1)\eta)\right]^{2N-1} \,, \non\\
c_{2} &=& \frac{\cosh((6n+5)\eta)}{2 \sinh((4n+2)\eta)\, 
\cosh((2n+3)\eta)} + \frac{i \cosh(2 n \eta)}{\cosh \eta + i \sinh(2 
n \eta)}\,.
\label{HIItc1c2}
\ee
The Hamiltonian (\ref{HamiltonianII}) is not Hermitian for either 
real or imaginary values of $\eta$.

\section{Quantum group symmetries}\label{sec:qg}

We first review the $U_{q}(B_{n})$ symmetry of 
the Hamiltonian corresponding to the first set of
K-matrices (\ref{KsetI}).  We then argue that the Hamiltonian corresponding to
the second set of K-matrices (\ref{KsetII}) has the quantum group
symmetry $U_{q}(C_{n})$. 

\subsection{Set I: $U_{q}(B_{n})$ symmetry}\label{sec:Bnsym}

It was already argued in \cite{Mezincescu:1990ui} that the 
Hamiltonian ${\cal H}^{(I)}$ (\ref{HamiltonianI})
corresponding to the first set of K-matrices (\ref{KsetI}) has $U_{q}(B_{n})$
symmetry.  It was subsequently shown in \cite{Mezincescu:1991rb} 
(generalizing the arguments in \cite{Kulish:1991np} for the XXZ chain)
that this symmetry extends to the full transfer matrix $t(u)$ 
(\ref{transfer}). Here we 
explicitly construct the coproduct of the generators, and show that 
they commute with the Hamiltonian.

For the vector representation of $B_{n} = O(2n+1)$, in the so-called {\em orthogonal} basis, 
the Cartan generators $\{ H_{1}, \ldots, H_{n}\}$ are given by 
the diagonal matrices \footnote{Explicit matrix representations for the generators can be 
obtained from e.g. \cite{Fonseca:2011sy} or Maple.}
\be
H_{\alpha} = e_{\alpha, \alpha} - e_{2n+2-\alpha, 2n+2-\alpha}\,, \qquad \alpha = 1,2, \ldots, n \,,
\label{CartanBn}
\ee
and the generators $\{ E^{\pm}_{1}, \ldots, E^{\pm}_{n}\}$
corresponding to the simple roots are given by 
\be
E^{+}_{\alpha} = e_{\alpha,\alpha+1}  +  e_{2n+1-\alpha, 
2n+2-\alpha}\,, \qquad E^{-}_{\alpha} = E^{+\, t}_{\alpha}\,, \qquad
\alpha = 1,2, \ldots, n \,.
\label{EpmBn}
\ee 
Indeed, these generators satisfy
\be
\left[ H_{i} \,,  E^{\pm}_{j} \right] = \pm \alpha_{i}^{(j)} 
E^{\pm}_{j}\,, \qquad i, j = 1,2, \ldots, n \,,
\ee
where $\{\alpha^{(1)}, \ldots, \alpha^{(n)}\}$ are the simple roots 
of $B_{n}$ in the orthogonal basis (see e.g. 
\cite{Feger:2012bs})
\be
\alpha^{(1)} &=& (1,-1, 0, \ldots, 0) \,, \non\\
\alpha^{(2)} &=& (0, 1,-1, 0, \ldots, 0) \,, \non\\
&\vdots& \non\\
\alpha^{(n-1)} &=& (0, \ldots, 0, 1,-1) \,, \non\\
\alpha^{(n)} &=& (0, \ldots, 0, 1) \,. 
\label{simplerootsBn}
\ee

Let us define the following coproduct for these generators
\be
\Delta(H_{j}) &=& H_{j} \otimes \id + \id \otimes H_{j} \,, \non \\
\Delta(E^{\pm}_{j}) &=& E^{\pm}_{j} \otimes e^{i \pi H_{j}} e^{\eta 
(H_{j} - H_{j+1})} + e^{-i \pi H_{j}} e^{-\eta 
(H_{j} - H_{j+1})} \otimes E^{\pm}_{j} \,, 
\label{coproductBn}
\ee
where $j = 1, \ldots, n$ with $H_{n+1} \equiv 0$. We observe that
\be
\Omega_{ij}\, \Delta(E^{+}_{i})\,  \Delta(E^{-}_{j}) -  \Delta(E^{-}_{j})\, 
\Delta(E^{+}_{i})\, \Omega_{ij} = \delta_{i,j} \frac{q^{ 
\Delta(H_{i})-\Delta(H_{i+1})}-q^{-\Delta(H_{i})+\Delta(H_{i+1})}}{q-q^{-1}}\,,
\ee
where $q=e^{2\eta}$ and
\be
\Omega_{ij} = \left\{ \begin{array}{cc}
e^{i \pi H_{{\rm max}(i,j)}}\otimes \id & \qquad |i-j| = 1 \\
\id \otimes \id  & \qquad |i-j| \ne 1 
\end{array} \right. \,.
\label{Omega}
\ee

The two-site Hamiltonian (\ref{twosite}) commutes
with the coproducts (\ref{coproductBn})
\be
\left[ \Delta(H_{j}) \,, h_{1,2}  \right] = \left[ \Delta(E^{\pm}_{j}) 
\,, h_{1,2} \right] = 0 \,, \qquad j = 1, \ldots, n \,.
\ee 
Since the $N$-site Hamiltonian is given (\ref{HamiltonianI}) by the 
sum of two-site Hamiltonians, it follows that 
the $N$-site Hamiltonian commutes with the $N$-fold coproducts
\be
\left[ \Delta_{(N)}(H_{j}) \,, {\cal H}^{(I)} \right] = \left[ \Delta_{(N)}(E^{\pm}_{j}) 
\,, {\cal H}^{(I)} \right] = 0 \,, \qquad j = 1, \ldots, n \,.
\label{UqBnsymmetry}
\ee 
This provides an explicit demonstration of the $U_{q}(B_{n})$ 
invariance of the Hamiltonian ${\cal H}^{(I)}$.

\subsubsection{Degeneracies and multiplicities for $U_{q}(B_{n})$}\label{sec:degmultI}

One of the important consequences of the $U_{q}(B_{n})$ symmetry of
the Hamiltonian is that the energy eigenstates form irreducible
representations of this algebra.  For generic values of $\eta$ (i.e.,
$\eta \ne i \pi/p$, where $p$ is a rational number), the
representations are the same as for the classical algebra $B_{n}$. The generalization of the 
familiar Clebsch-Gordan theorem from $A_{1} = SU(2)$ to $B_{n}$ 
implies that the $N$-site Hilbert space has a decomposition of the 
form 
\be
{\cal V}^{(2n+1) \otimes N} = \bigoplus_{j} d^{(j, N, n)}\, {\cal V}^{(j)} \,,
\label{decomBn}
\ee
where ${\cal V}^{(j)}$ denotes an irreducible representation of $B_{n}$
with dimension $j$ (= degeneracy of the corresponding 
energy eigenvalue) and $d^{(j, N, n)}$ is its multiplicity.  
Here we specify the irreducible representations by their dimensions, 
and we allow for the possibility that there can be more than one inequivalent 
irreducible representation with a given dimension. For example, 
$B_{2}$ has a ${\bf 35}$ and a ${\bf 35'}$.

The first few cases are as follows (see e.g. \cite{Feger:2012bs}): 
\footnote{For later reference, we also present the tensor-product decompositions in 
terms of the Dynkin labels $[a_{1}, \ldots, a_{n}]$ of the representations.}
\be
B_{1}:\qquad  N&=&2: \qquad {\bf 3} \otimes  {\bf 3} = {\bf 1} \oplus  {\bf 3} 
\oplus  {\bf 5} \non \\
& &  \qquad\qquad\qquad = [0] \oplus [2] \oplus [4] \non\\
 N&=&3:\qquad {\bf 3} \otimes  {\bf 3} \otimes  {\bf 3} = {\bf 1} 
 \oplus  3 \cdot {\bf 3} 
\oplus 2 \cdot {\bf 5} \oplus  {\bf 7} \non\\
& &  \qquad\qquad\qquad\qquad = [0] \oplus  3 [2] \oplus  2 [4] \oplus  [6]
\label{decompB1}
\ee 

\be
B_{2}:\qquad  N&=&2: \qquad {\bf 5} \otimes  {\bf 5} = {\bf 1} 
\oplus  {\bf 10} 
\oplus  {\bf 14} \non \\
& &  \qquad\qquad\qquad = [0,0] \oplus [0,2] \oplus [2,0] \non\\
 N&=&3:\qquad {\bf 5} \otimes  {\bf 5} \otimes  {\bf 5} = 3 \cdot 
 {\bf 5} 
 \oplus  {\bf 10} 
\oplus {\bf 30} \oplus   2 \cdot {\bf 35} \non \\
& &  \qquad\qquad\qquad\qquad = 3 [1,0] \oplus [0,2] \oplus [3,0] 
\oplus 2 [1,2] 
\label{decomp2}
\ee 

\be
B_{3}:\qquad  N&=&2: \qquad {\bf 7} \otimes  {\bf 7} = {\bf 1} 
\oplus  {\bf 21} 
\oplus  {\bf 27} \non \\
& &  \qquad\qquad\qquad = [0,0,0] \oplus [0,1,0] \oplus  [2,0,0] 
\non\\
 N&=&3:\qquad {\bf 7} \otimes  {\bf 7} \otimes  {\bf 7} = 3 \cdot 
 {\bf 7} 
 \oplus  {\bf 35} 
\oplus {\bf 77} \oplus   2 \cdot {\bf 105}\non \\
& &  \qquad\qquad\qquad\qquad =  3 [1,0,0] \oplus  [0,0,2] \oplus  
[3,0,0] \oplus 2 [1,1,0] 
\label{decompB3}
\ee 

We have verified numerically that the Hamiltonian as well as the
transfer matrix for set I  (\ref{KsetI}) have exactly these degeneracies and multiplicities 
for generic values of $\eta$, which provides further evidence of 
their $U_{q}(B_{n})$ invariance.

\subsection{Set II: $U_{q}(C_{n})$ symmetry}\label{sec:Cnsym}

For the vector representation of $C_{n} = Sp(2n)$ in the orthogonal basis, the 
Cartan generators are given by
\be
\tilde H_{\alpha} = \tilde e_{\alpha, \alpha} - \tilde 
e_{2n+1-\alpha, 2n+1-\alpha}\,, \qquad \alpha = 1,2, \ldots, n \,,
\label{Cngens1}
\ee
and the generators corresponding to the simple roots are given by
\be
\tilde E^{+}_{\alpha} &=& \tilde e_{\alpha,\alpha+1}  +  \tilde e_{2n-\alpha, 
2n+1-\alpha}\,, \qquad
\alpha = 1,2, \ldots, n-1 \,, \non \\
\tilde E^{+}_{n} &=&  \tilde e_{n, n+1} \,,
\label{Cngens2}
\ee 
and $\tilde E^{-}_{\alpha} = \tilde E^{+\, t}_{\alpha}$, where 
$\tilde e_{\alpha \beta}$ are the elementary $(2n) \times (2n)$ 
matrices. These generators satisfy
\be
\left[ \tilde H_{i} \,,  \tilde E^{\pm}_{j} \right] = \pm \alpha_{i}^{(j)} 
\tilde E^{\pm}_{j}\,, \qquad i, j = 1,2, \ldots, n \,,
\ee
where $\{\alpha^{(1)}, \ldots, \alpha^{(n)}\}$ are the simple roots 
of $C_{n}$ in the orthogonal basis
\be
\alpha^{(1)} &=& (1,-1, 0, \ldots, 0) \,, \non\\
\alpha^{(2)} &=& (0, 1,-1, 0, \ldots, 0) \,, \non\\
&\vdots& \non\\
\alpha^{(n-1)} &=& (0, \ldots, 0, 1,-1) \,, \non\\
\alpha^{(n)} &=& (0, \ldots, 0, 2) \,,
\label{simplerootsCn}
\ee
c.f. (\ref{simplerootsBn}).

Let us now consider the Hamiltonian ${\cal H}^{(II)}$ (\ref{HamiltonianII}) corresponding to the second set of 
K-matrices (\ref{KsetII}). 
The appearance of $U_{q}(C_{n})$ symmetry in this spin chain can be understood as a sort 
of ``breaking'' of $B_{n}$ down to $C_{n}$. That is,
we consider an embedding of $C_{n}$ in $B_{n}$, such that the vector 
space ${\cal V}^{(2n+1)}$ at each site, which forms a 
$(2n+1)$-dimensional irreducible representation of 
$B_{n}$, decomposes into the direct sum of the $2n$-dimensional and 
1-dimensional irreducible representations of $C_{n}$,
\be
{\cal V}^{(2n+1)} = {\cal W}^{(2n)} \oplus {\cal W}^{(1)} \,.
\ee 
We construct the corresponding generators of $C_{n}$ on ${\cal 
V}^{(2n+1)}$ by starting from the 
vector representation of the $C_{n}$ generators
in terms of $(2n) \times (2n)$ matrices (\ref{Cngens1})-(\ref{Cngens2}), 
and then inserting a column of 0's between columns $n$ and $n+1$, 
and a row of 0's between rows $n$ and $n+1$, thereby arriving at a 
set of $(2n+1) \times (2n+1)$ matrices. That is,
\be
\left(\begin{array}{cc}
A & B\\
C & D 
\end{array}\right) \mapsto \left(\begin{array}{ccc}
A & 0 & B\\
0 & 0 & 0\\
C & 0 & D 
\end{array}\right) \,,
\label{embedding}
\ee
where $A, B, C, D$ represent $n \times n$ matrices.

In short, we henceforth represent the generators of $C_{n}$ by $(2n+1) 
\times (2n+1)$ matrices, such that the Cartan generators are given by 
the diagonal matrices
\be
H_{\alpha} = e_{\alpha, \alpha} - e_{2n+2-\alpha, 2n+2-\alpha}\,, \qquad \alpha = 1,2, \ldots, n \,,
\label{CartanCn}
\ee
and the generators corresponding to the simple roots are given by 
\be
E^{+}_{\alpha} &=& e_{\alpha,\alpha+1}  +  e_{2n+1-\alpha, 
2n+2-\alpha}\,, \qquad \alpha = 1,2, \ldots, n-1 \,, \non\\
E^{+}_{n} &=& e_{n,n+2} \,,
\label{EpmCn}
\ee 
and $E^{-}_{\alpha} = E^{+\, t}_{\alpha}$. Comparing with the 
corresponding expressions for the generators of $B_{n}$ (\ref{CartanBn})-(\ref{EpmBn}), 
we see that they are exactly the same, except for $E^{\pm}_{n}$.

Below we shall also need another pair of generators of $C_{n}$, which we denote by 
$E^{\pm}_{0}$ 
\be
E^{+}_{0} = e_{1,2n+1} \,, \qquad E^{-}_{0} = e_{2n+1,1} \,,
\label{E0}
\ee
which are related to $E^{\pm}_{n}$ as follows
\be
E^{\pm}_{n} =\left\{\begin{array}{cl}
E^{\pm}_{0} & n=1 \\
-\tfrac{1}{2} [[ E^{\pm}_{0}\,, E_{1}^{\mp}]\,, E_{1}^{\mp}] & n=2 \\
\tfrac{1}{4}[[[[ E^{\pm}_{0}\,, E_{1}^{\mp}]\,, E_{1}^{\mp}]\,, E_{2}^{\mp}] \,, E_{2}^{\mp}] & n=3 \\
\vdots & \quad\vdots \\
(-\tfrac{1}{2})^{n-1}[[\ldots [[ E^{\pm}_{0}\,, E_{1}^{\mp}]\,, 
E_{1}^{\mp}]\,, \ldots E_{n-1}^{\mp}] \,, E_{n-1}^{\mp}]
& \quad n 
\end{array} \right. \,,
\label{Enested}
\ee
where the final line has a $2(n-1)$-fold multiple commutator.

The Cartan generators have the usual coproduct
\be
\Delta(H_{j}) = H_{j} \otimes \id + \id \otimes H_{j} \,, \qquad j = 
1, \ldots, n\,,
\label{coproductCnH}
\ee
and we propose the following coproducts for the first $n-1$ 
raising/lowering operators
\be
\Delta(E^{\pm}_{j}) &=& E^{\pm}_{j} \otimes e^{i \pi H_{j+1}} + e^{i 
\pi H_{j+1}} e^{-2\eta 
(H_{j} - H_{j+1})} \otimes E^{\pm}_{j} \,, \qquad j = 
1, \ldots, n-1\,.
\label{coproductCn}
\ee
We have not succeeded to find such a simple expression for the coproduct for $E^{\pm}_{n}$. 
However, we observe that the generators $E^{\pm}_{0}$ (\ref{E0}) 
do have a simple coproduct
\be
\Delta(E^{\pm}_{0}) = E^{\pm}_{0} \otimes \id +  e^{4\eta H_{1}} 
\otimes E^{\pm}_{0} \,.
\ee
Hence, using (\ref{Enested}), we obtain the result
\be 
\Delta(E^{\pm}_{n}) = (-\tfrac{1}{2})^{n-1}[[\ldots [[ 
\Delta(E^{\pm}_{0})\,, \Delta(E_{1}^{\mp})]\,, 
\Delta(E_{1}^{\mp})]\,, \ldots \Delta(E_{n-1}^{\mp})] \,, 
\Delta(E_{n-1}^{\mp})] \,.
\label{coproductCnlast}
\ee
These expressions for the coproducts satisfy the 
coassociativity property \cite{Chari:1994pz} \footnote{An earlier version of this paper 
had a different expression for $\Delta(E^{\pm}_{n})$, which did not 
satisfy the coassociativity property.}
\be
(\Delta \otimes \id)  \Delta = (\id \otimes \Delta) \Delta \,.
\ee

We observe the following relations for $1 
\le i , j < n$ : 
\be
\Delta(E^{+}_{i})\,  \Delta(E^{-}_{i}) - e^{4\eta}\, 
\Delta(E^{-}_{i})\,  \Delta(E^{+}_{i}) &=& 
\frac{e^{-4\eta(\Delta(H_{i})-\Delta(H_{i+1}))}- 
\id\otimes\id}{e^{-4\eta}-1} \,, \non \\
e^{2\eta}\, \Omega_{ij}\, \Delta(E^{+}_{i})\,  \Delta(E^{-}_{j})  &=&
\Delta(E^{-}_{j})\, \Delta(E^{+}_{i})\, \Omega_{ij} \,, \qquad   
|i-j| = 1 \,, \label{EpiEmj} \\
\Delta(E^{+}_{i})\,  \Delta(E^{-}_{j})  &=& \Delta(E^{-}_{j})\, \Delta(E^{+}_{i})   
\,, \qquad \quad |i-j| \ge 2 \,, \non 
\ee
where $\Omega_{ij}$ is given by (\ref{Omega}).

By construction, the coproducts 
(\ref{coproductCnH})-(\ref{coproductCnlast})
commute with the ``new'' two-site Hamiltonian (\ref{twositeII})
\be
\left[ \Delta(H_{j}) \,, \tilde h_{1,2}  \right] = \left[ \Delta(E^{\pm}_{j}) 
\,, \tilde h_{1,2} \right] = 0 \,, \qquad j = 1, \ldots, n \,.
\ee 
Moreover, all the generators (whose row $(n+1)$ and column $(n+1)$ 
are null, as in (\ref{embedding})) evidently commute with $U=e_{n+1,n+1}$. Since the $N$-site Hamiltonian is 
given (\ref{HamiltonianII}) by the 
sum of two-site Hamiltonians and a term proportional to $U_{N}$, it follows that 
the $N$-site Hamiltonian commutes with the $N$-fold coproducts
\be
\left[ \Delta_{(N)}(H_{j}) \,, {\cal H}^{(II)} \right] = \left[ \Delta_{(N)}(E^{\pm}_{j}) 
\,, {\cal H}^{(II)} \right] = 0 \,, \qquad j = 1, \ldots, n \,,
\label{UqCnsymmetry}
\ee 
which implies the $U_{q}(C_{n})$ invariance of the Hamiltonian ${\cal 
H}^{(II)}$. We conjecture that this symmetry also extends to the full 
transfer matrix.
The symmetry for the case $n=1$ (note that $C_{1} = A_{1}$) was first noted in \cite{Nepomechie:1999jz}.

\subsubsection{Degeneracies and multiplicities for $U_{q}(C_{n})$}\label{sec:degmultII}

The $U_{q}(C_{n})$ invariance of the Hamiltonian implies that,
for generic values of $\eta$,  the $N$-site Hilbert space 
has a decomposition of the form (cf. Eq. (\ref{decomBn}))
\be
\left( {\cal W}^{(2n)} \oplus {\cal W}^{(1)} \right)^{\otimes N} = 
\bigoplus_{j} \tilde d^{(j, N, n)}\, {\cal W}^{(j)} \,,
\label{decomCn}
\ee
where ${\cal W}^{(j)}$ denotes an irreducible representation of $C_{n}$
with dimension $j$ (= degeneracy of the corresponding 
energy eigenvalue) and $\tilde d^{(j, N, n)}$ is its multiplicity.

The first few cases are as follows (see again e.g. \cite{Feger:2012bs}):
\be
C_{1} = A_{1}:\qquad  N&=&2: \qquad \left({\bf 2} \oplus {\bf 
1}\right)^{\otimes 2} 
= 2\cdot {\bf 1} \oplus  2 \cdot{\bf 2} 
\oplus  {\bf 3} \non \\
& &  \qquad\qquad\qquad\qquad = 2 [0] \oplus 2 [1] \oplus [2] \non\\
 N&=&3:\qquad \left({\bf 2} \oplus {\bf 1}\right)^{\otimes 3} 
 = 4 \cdot {\bf 1} 
 \oplus  5 \cdot {\bf 2} 
\oplus 3 \cdot {\bf 3} \oplus  {\bf 4} \non\\
& &  \qquad\qquad\qquad\qquad = 4 [0] \oplus 5 [1] \oplus 3 [2] 
\oplus [3]
\label{decompC1}
\ee 
 
\be
C_{2}:\qquad  N&=&2: \qquad \left({\bf 4} \oplus {\bf 1}\right)^{\otimes 2} 
= 2\cdot {\bf 1} \oplus  2 \cdot{\bf 4} 
\oplus  {\bf 5} \oplus  {\bf 10} \non \\
& &  \qquad\qquad\qquad\qquad = 2 [0,0] \oplus 2 [1,0] \oplus [0,1] \oplus 
[2,0] \non\\
 N&=&3:\qquad \left({\bf 4} \oplus {\bf 1}\right)^{\otimes 3} 
 = 4 \cdot {\bf 1} 
 \oplus  6 \cdot {\bf 4} 
\oplus 3 \cdot {\bf 5} \oplus 3 \cdot {\bf 10} \oplus 2 \cdot {\bf 
16} \oplus {\bf 20} \non \\
& &  \qquad\qquad\qquad\qquad = 4 [0,0] \oplus 6 [1,0] \oplus 3 [0,1] 
\oplus 3 [2,0] \oplus 2 [1,1] \oplus [3,0] \non\\
\label{decompC2}
\ee 

\be
C_{3}:\qquad  N&=&2: \qquad \left({\bf 6} \oplus {\bf 1}\right)^{\otimes 2} 
= 2\cdot {\bf 1} \oplus  2 \cdot{\bf 6} 
\oplus  {\bf 14} \oplus  {\bf 21} \non \\
& &  \qquad\qquad\qquad\qquad = 2 [0,0,0] \oplus 2 [1,0,0] \oplus 
[0,1,0] \oplus [2,0,0] \non\\
 N&=&3:\qquad \left({\bf 6} \oplus {\bf 1}\right)^{\otimes 3} 
 = 4 \cdot {\bf 1} 
 \oplus  6 \cdot {\bf 6} 
\oplus 3 \cdot {\bf 14} \oplus {\bf 14'} \oplus 3 \cdot {\bf 21} \oplus {\bf 
56} \oplus  2 \cdot {\bf 64}\non \\
& &  \qquad\qquad\qquad\qquad = 4 [0,0,0] \oplus 6 [1,0,0] \oplus 3 
[0,1,0] \oplus [0,0,1] \oplus \non\\
& & \qquad\qquad\qquad\qquad\qquad \oplus 3 [2,0,0] \oplus [3,0,0] \oplus 2 
[1,1,0] 
\label{decompC3}
\ee 

We have verified numerically that the Hamiltonian as well as the
transfer matrix for set II  (\ref{KsetII}) have exactly these degeneracies and multiplicities 
for generic values of $\eta$, which provides further evidence of 
their $U_{q}(C_{n})$ invariance.

\section{Bethe ansatz}\label{sec:BA}

Our discussion so far has not made use of the integrability of the 
models. However, this integrability has been exploited to obtain Bethe 
ansatz solutions of the models corresponding to sets I (\ref{KsetI}) 
and II  (\ref{KsetII}) in \cite{Artz:1994qy} and \cite{Li:2005pp}, 
respectively. \footnote{The solution of the $A_{2n}^{(2)}$ family of 
integrable quantum spin chains has a long history. The initial work was for {\em closed} chains with periodic 
boundary conditions. The case $n=1$ (corresponding to the Izergin-Korepin model 
\cite{Izergin:1980pe}) was first solved using the analytical Bethe 
ansatz approach \cite{Vichirko1983, Reshetikhin:1983vw}, which gave the eigenvalues (but not 
the eigenvectors) of the transfer matrix. This approach was subsequently extended to 
$n>1$ in \cite{Reshetikhin:1987}. The algebraic Bethe ansatz for the 
case $n=1$, which gave also the eigenvectors of the transfer matrix, 
was formulated in the important work \cite{Tarasov:1989}.
The seminal work of Sklyanin \cite{Sklyanin:1988yz} made it possible to 
generalize these results to {\em open} $A_{2n}^{(2)}$ chains.
The case $n=1$ with the first set of K-matrices (\ref{KsetI})
was solved using the analytical Bethe ansatz approach in 
\cite{Mezincescu:1991ag}, and this approach was subsequently extended to 
$n>1$ in \cite{Artz:1994qy}. The algebraic Bethe ansatz for the 
case $n=1$ was developed in \cite{Fan:1997wpk, Li:2003ucc}. Finally, 
the algebraic Bethe ansatz for $n>1$ with general diagonal K-matrices 
\cite{Batchelor:1996np, LimaSantos:2002ui} was formulated in 
\cite{Li:2005pp}. An analytical Bethe ansatz 
approach for the case $n=1$ with general non-diagonal K-matrices has 
recently been formulated  in \cite{Hao:2014fha}. Other related work 
includes \cite{Yung:1994td, Batchelor:1995gx, Kim:1994aha, Fan:1999, 
Lima-Santos:1999vph, Kurak:2004ip, Li:2006mv, Pimenta:2014} .}

Here we study how the quantum group symmetry of these models is
reflected in their Bethe ansatz solutions.  Our main result is a
formula for the Dynkin label $[a_{1}, \ldots, a_{n}]$ of a
Bethe state in terms of the cardinalities $(m_{1}, \ldots,
m_{n})$ of the corresponding Bethe roots (i.e., $m_{i}$ is the number 
of Bethe roots of type $i$, where $i=1, \ldots , n$), see Eq.  (\ref{DynkinBArltn}).  
The Dynkin label
uniquely characterizes an irreducible representation, and in particular
determines its dimension, which is the degeneracy of the
corresponding eigenvalue.  The number of distinct solutions of the
Bethe equations with $(m_{1}, \ldots, m_{n})$ Bethe roots
determines the multiplicity.  We shall then 
verify numerically in Sec. \ref{sec:completeness} that, in this way, 
the patterns 
of degeneracies and multiplicities predicted by the quantum group 
symmetry (\ref{decompB1})-(\ref{decompB3}) and  
(\ref{decompC1})-(\ref{decompC3}) are completely accounted for by the 
Bethe ansatz solutions.

\subsection{Review of the Bethe ansatz solutions}\label{sec:BAreview}

Before presenting our formula for the Dynkin labels, we briefly 
summarize here the Bethe ansatz solutions of the models.
The Bethe states, which we denote by 
\be
|\Lambda^{(m_{1}, \ldots, m_{n})} \rangle = |\{u^{(1)}_{1}, \ldots, 
u^{(1)}_{m_{1}}\}, \ldots, \{u^{(n)}_{1}, \ldots, 
u^{(n)}_{m_{n}}\} \rangle \,,
\label{Bethestates}
\ee
depend on $n$ sets of Bethe roots $\{u^{(1)}_{1}, \ldots, 
u^{(1)}_{m_{1}}\}, \ldots, \{u^{(n)}_{1}, \ldots, 
u^{(n)}_{m_{n}}\} $,
which are solutions of the following $n$ sets of Bethe equations 
\cite{Artz:1994qy, Li:2005pp}
\be
e_{1}^{2N}(u^{(1)}_{k}) &=& \prod_{j=1,\, j\ne k}^{m_{1}}
e_{2}(u^{(1)}_{k}-u^{(1)}_{j})\, e_{2}(u^{(1)}_{k}+u^{(1)}_{j})
\prod_{j=1}^{m_{2}}
e_{-1}(u^{(1)}_{k}-u^{(2)}_{j})\, e_{-1}(u^{(1)}_{k}+u^{(2)}_{j}) \,, 
\non \\
& & \qquad\qquad\qquad\qquad k = 1, \ldots, m_{1}\,, \non \\
1 &=& \prod_{j=1}^{m_{l-1}}
e_{-1}(u^{(l)}_{k}-u^{(l-1)}_{j})\, e_{-1}(u^{(l)}_{k}+u^{(l-1)}_{j})
\prod_{j=1,\, j\ne k}^{m_{l}}
e_{2}(u^{(l)}_{k}-u^{(l)}_{j})\, e_{2}(u^{(l)}_{k}+u^{(l)}_{j}) \non \\
& & \times
\prod_{j=1}^{m_{l+1}}
e_{-1}(u^{(l)}_{k}-u^{(l+1)}_{j})\, e_{-1}(u^{(l)}_{k}+u^{(l+1)}_{j})
\,, \quad k=1, \ldots, m_{l}\,, \quad l = 2, \ldots, n-1\,, \non \\
\chi(u^{(n)}_{k}) &=& \prod_{j=1}^{m_{n-1}}
e_{-1}(u^{(n)}_{k}-u^{(n-1)}_{j})\, 
e_{-1}(u^{(n)}_{k}+u^{(n-1)}_{j})\non \\
& & \times \prod_{j=1,\, j\ne k}^{m_{n}}
e_{2}(u^{(n)}_{k}-u^{(n)}_{j})\, e_{-1}(u^{(n)}_{k}-u^{(n)}_{j}+i\pi)\, 
e_{2}(u^{(n)}_{k}+u^{(n)}_{j})\, 
e_{-1}(u^{(n)}_{k}+u^{(n)}_{j}+i\pi)\,,\non \\
& & \qquad\qquad\qquad\qquad k = 1, \ldots, m_{n}\,,
\label{BAE}
\ee
where here we use the compact notation
\be
e_{k}(u) = \frac{\sinh(\frac{u}{2} +\eta\, k)}{\sinh(\frac{u}{2} 
-\eta\, k)} \,,
\ee
and
\be
\chi(u) = \left\{ \begin{array}{cr}
1 & \mbox{ for } U_{q}(B_{n})\\
\left(\frac{\sinh(\frac{1}{2}(u +\eta-\frac{i \pi}{2}))}
{\sinh(\frac{1}{2}(u -\eta+\frac{i \pi}{2}))}\right)^{2} &  \mbox{ 
for } U_{q}(C_{n})
\end{array}\right. \,.
\ee
The above equations are for $n>1$. For $n=1$, the Bethe equations are 
given by
\be
e_{1}^{2N}(u^{(1)}_{k})\, \chi(u^{(1)}_{k})  &=& \prod_{j=1,\, j\ne k}^{m_{1}}
e_{2}(u^{(1)}_{k}-u^{(1)}_{j})\, e_{-1}(u^{(1)}_{k}-u^{(1)}_{j}+i\pi) 
\non \\
& & \qquad \times
e_{2}(u^{(1)}_{k}+u^{(1)}_{j})\, 
e_{-1}(u^{(1)}_{k}+u^{(1)}_{j}+i\pi)\,, \quad k = 1, \ldots, m_{1} \,.
\ee

The Bethe states are certain simultaneous eigenstates of the transfer matrix $t(u)$ (\ref{transfer})
and the Cartan generators $\Delta_{(N)}(H_{i})$ (\ref{CartanBn}), (\ref{coproductBn}), 
(\ref{CartanCn}), (\ref{coproductCnH}),
\be
t(u)\, |\Lambda^{(m_{1}, \ldots, m_{n})} \rangle &=& \Lambda^{(m_{1}, 
\ldots, m_{n})}(u)\, |\Lambda^{(m_{1}, \ldots, m_{n})} \rangle \,, 
\non \\
\Delta_{(N)}(H_{i})\, |\Lambda^{(m_{1}, \ldots, m_{n})} \rangle &=& 
h_{i}\, |\Lambda^{(m_{1}, \ldots, m_{n})} \rangle \,, \qquad i = 1, 
\ldots, n \,.
\label{simultan}
\ee 
The eigenvalues of the transfer matrix are given by \cite{Artz:1994qy, Li:2005pp}
\begin{IEEEeqnarray}{rCl}
& & \Lambda^{(m_1 \,, \cdots \,, m_n)}(u) \non \\
& &  =  A^{(m_1)}(u)\, \psi_{1}(u)\
\frac{\sinh(u-2(2n+1)\eta)}{\sinh(u-2\eta)}
\frac{\cosh(u-(2n-1)\eta)}{\cosh(u-(2n+1)\eta)} 
\left[ 2 \sinh (\uh -2\eta) \cosh(\uh - (2n+1)\eta) \right]^{2N} \non \\
&  & + C^{(m_1)}(u)\, \tilde \psi_{1}(u)\
\frac{\sinh u}{\sinh (u-4n\eta)}
\frac{\cosh(u-(2n+3)\eta)}{\cosh(u-(2n+1)\eta)} 
\left[ 2 \sinh (\uh) \cosh(\uh - (2n-1)\eta)\right]^{2N} \non \\
&  & + \left\{ w(u)\, \psi_{2}(u)\, B_n^{(m_n)}(u) +
\sum_{l=1}^{n-1} \left[ z_l(u)\, \psi_{1}(u)\, B_l^{(m_l \,, m_{l+1})}(u)
+ \tilde z_l(u)\, \tilde \psi_{1}(u)\, \tilde B_l^{(m_l \,, m_{l+1})}(u) \right] \right\}  \non\\
& &\qquad \times \left[2 \sinh(\uh) \cosh(\uh -(2n+1)\eta)\right]^{2N}\,,
\label{Lambda}
\end{IEEEeqnarray}
where
\be
A^{(m_1)}(u)&=&\prod_{j=1}^{m_1}
\frac{\sinh({1\over 2}({u-u_j^{(1)}})+\eta)\
      \sinh({1\over 2}({u+u_j^{(1)}})+\eta)}
     {\sinh({1\over 2}({u-u_j^{(1)}})-\eta)\
      \sinh({1\over 2}({u+u_j^{(1)}})-\eta)}\,,
\ee
\be
C^{(m_1)}(u)&=&A^{(m_1)}(-u-\rho)  \non \\
&=&\prod_{j=1}^{m_1}
\frac{\cosh({1\over 2}({u-u_j^{(1)}})-2(n+1)\eta)\
      \cosh({1\over 2}({u+u_j^{(1)}})-2(n+1)\eta)}
     {\cosh({1\over 2}({u-u_j^{(1)}})-2n\eta)\
      \cosh({1\over 2}({u+u_j^{(1)}})-2n\eta)}\,, 
\ee
\be
B_l^{(m_l \,, m_{l+1})}(u)&=& \prod_{j=1}^{m_l}
\frac{\sinh({1\over 2}({u-u_j^{(l)}})-(l+2)\eta)\
      \sinh({1\over 2}({u+u_j^{(l)}})-(l+2)\eta)}
     {\sinh({1\over 2}({u-u_j^{(l)}})-l\eta)\
      \sinh({1\over 2}({u+u_j^{(l)}})-l\eta)} \non\\
&\times & \prod_{j=1}^{m_{l+1}}
\frac{\sinh({1\over 2}({u-u_j^{(l+1)}})-(l-1)\eta)\
      \sinh({1\over 2}({u+u_j^{(l+1)}})-(l-1)\eta)}
     {\sinh({1\over 2}({u-u_j^{(l+1)}})-(l+1)\eta)\
      \sinh({1\over 2}({u+u_j^{(l+1)}})-(l+1)\eta)} \non \\
\tilde B_l^{(m_l \,, m_{l+1})}(u)&=& B_l^{(m_l \,, m_{l+1})}(-u-\rho) \,,
 \qquad l=1\,, \cdots \,, n-1 \,,  \label{A-B} 
\ee
\be
B_n^{(m_n)}(u)&=&\prod_{j=1}^{m_n}
\frac{\sinh({1\over 2}({u-u_j^{(n)}})-(n+2)\eta)\
      \sinh({1\over 2}({u+u_j^{(n)}})-(n+2)\eta)}
     {\sinh({1\over 2}({u-u_j^{(n)}})-n\eta)\
      \sinh({1\over 2}({u+u_j^{(n)}})-n\eta)}\non\\
& & \times
\frac{\cosh({1\over 2}({u-u_j^{(n)}})-(n-1)\eta)\
      \cosh({1\over 2}({u+u_j^{(n)}})-(n-1)\eta)}
     {\cosh({1\over 2}({u-u_j^{(n)}})-(n+1)\eta)\
      \cosh({1\over 2}({u+u_j^{(n)}})-(n+1)\eta)} \,, 
\ee
and
\be
z_l(u) & = &  \frac{\sinh(u)}{\sinh(u-2l\eta)}
\frac{\sinh(u-2(2n+1)\eta)}{\sinh(u-2(l+1)\eta)}
\frac{\cosh(u-(2n-1)\eta)}{\cosh(u-(2n+1)\eta)} \,,  \non \\
\tilde z_l(u) &=& z_l(-u -\rho)                 \,, 
\qquad\qquad\qquad l = 1 \,, \cdots \,, n-1\,,  \non \\
w(u) & = & \frac{\sinh(u)}{\sinh(u-2n\eta)}
\frac{\sinh(u-2(2n+1)\eta)}{\sinh(u-2(n+1)\eta)} \,,
\label{z-w}
\ee
where
\be
\psi_{1}(u) &=& \left\{ \begin{array}{cr}
1 & \mbox{ for } U_{q}(B_{n})\\
\frac{\cosh(u -(2n+3)\eta)}
{\cosh(u -(2n-1)\eta)}\left[\cosh \eta - i \sinh(u-2 n 
\eta)\right]^{2} &  \mbox{ for } U_{q}(C_{n})
\end{array}\right. \,, \non\\
\tilde \psi_{1}(u) &=& \psi_{1}(-u-\rho) \,, \non\\
\psi_{2}(u) &=& \left\{ \begin{array}{cr}
1 & \mbox{ for } U_{q}(B_{n})\\
\cosh(u -(2n+3)\eta)\, \cosh(u -(2n-1)\eta) &  \mbox{ for } U_{q}(C_{n})
\end{array}\right. \,.
\label{psi12}
\ee

The eigenvalues of both Hamiltonians ${\cal H}^{(I)}$ and ${\cal H}^{(II)}$ are given 
by
\be
E =-\sum_{k=1}^{m_{1}}\frac{\sinh(2\eta)}{2 \sinh(\frac{1}{2} u^{(1)}_{k}-\eta) 
\sinh(\frac{1}{2} u^{(1)}_{k}+\eta)} - 
\frac{(N-1)\cosh((2n+3)\eta)}{2\sinh(2\eta)\, \cosh((2n+1)\eta)} \,,
\label{BAenergy}
\ee
as follows from (\ref{HIt})-(\ref{HItc1c2}), (\ref{HIIt})-(\ref{HIItc1c2}), and (\ref{Lambda})-(\ref{psi12}).

The Bethe states have been constructed in \cite{Li:2005pp} using the
nested algebraic Bethe ansatz approach. The 
``double-row'' monodromy matrix 
\be
{\cal T}_{a}(u) = T_{a}(u)\, K^{-}_{a}(u)\, \hat T_{a}(u)
\label{scriptT}
\ee 
can be written as a $(2n+1) \times (2n+1)$ matrix in the auxiliary space whose 
matrix elements are operators on the quantum space ${\cal V}^{\otimes 
N}$
\be
{\cal T}_{a}(u) = \left( \begin{array}{cccccc}
A_{1}(u)  & B_{2}(u) & B_{3}(u) & \dots  & B_{2n}(u) & F(u) \\
 *        &   *      &   *      & \dots  &  *     &   *  \\
 \vdots   & \vdots   &  \vdots  & \cdots  & \vdots &   \vdots  \\
 *        &   *      &   *      & \dots  &  *     &   *  \\
G(u)      & C_{2}(u) & C_{3}(u) & \dots  & C_{2n}(u) & A_{2n+1}(u)
\end{array} \right)_{(2n+1) \times (2n+1)}\,.
\label{doublerowmonodromy}
\ee
The basic idea is to construct the Bethe states
using the $B_{i}(u)$ operators (as well as others) as creation 
operators acting on the reference state
\be
|0\rangle = \left(\begin{array}{c}
1 \\
0 \\
\vdots \\
0
\end{array}\right)_{2n+1}^{\otimes N} \,.
\label{reference}
\ee

We conjecture that the (on-shell) Bethe states are highest-weight states of the 
quantum group
\be
\Delta_{(N)}(E^{+}_{i})\, |\Lambda^{(m_{1}, \ldots, m_{n})} \rangle = 
0 \,, \qquad i = 1, \ldots, n \,,
\label{highestweight}
\ee
as is the case for other integrable open quantum spin chains with quantum group 
symmetry (see e.g. \cite{Pasquier:1989kd, Mezincescu:1991rb, 
Mezincescu:1991ag, Destri:1991zm, Destri:1992fa, devega:1994hf, 
Artz:1994qy, Foerster:1993fp, GonzalezRuiz:1994tw}). 
However, a proof of this conjecture is beyond the scope of this paper.
As a consequence of (\ref{highestweight}), degenerate 
eigenvectors  (i.e., linearly independent eigenvectors of the 
transfer matrix $t(u)$ whose corresponding eigenvalues coincide with the
eigenvalue $\Lambda^{(m_{1}, \ldots, m_{n})}(u)$
of the Bethe state $|\Lambda^{(m_{1}, \ldots, m_{n})} \rangle$) which are obtained by acting on the Bethe state with the lowering 
operators $\Delta_{(N)}(E^{-}_{i})$ form an 
irreducible representation of the algebra that is uniquely 
characterized by the (highest) weights of the Bethe state,
known as the Dynkin label.

\subsection{Dynkin labels of the Bethe states}\label{sec:Dynkin}

We propose that the Dynkin label $[a_{1}, \ldots, a_{n}]$ 
corresponding to a Bethe state
$|\Lambda^{(m_{1}, \ldots, m_{n})} \rangle$ whose Bethe roots have cardinalities $(m_{1},\ldots, m_{n})$ 
is given for $n>1$ by
\be
a_{1} &=& N - 2m_{1} + m_{2} \,, \non \\
a_{i} &=& m_{i-1} - 2m_{i} + m_{i+1} \,, \qquad i = 2, \ldots, n-1 
\,, \non \\
a_{n} &=& \left\{ \begin{array}{rr}
2 (m_{n-1} - m_{n}) & \mbox{ for } U_{q}(B_{n})\\
m_{n-1} - m_{n} &  \mbox{ for } U_{q}(C_{n})
\end{array}\right. \,.
\label{DynkinBArltn}
\ee
For $n=1$,
\be
a_{1} &=& \left\{ \begin{array}{rr}
2 (N - m_{1}) & \mbox{ for } U_{q}(B_{1})\\
N - m_{1} &  \mbox{ for } U_{q}(C_{1})
\end{array}\right. \,.
\label{DynkinBArltn2}
\ee 

It is convenient to divide the proof of this result into two parts. The first part of the proof is the relation of the 
eigenvalues $(h_{1}, \ldots, h_{n})$ of the Cartan generators to the cardinalities 
$(m_{1}, \ldots, m_{n})$ of the Bethe roots
\be
h_{1} &=& N - m_{1} \,, \non \\
h_{i} &=& m_{i-1} - m_{i} \,, \qquad i = 2, 3, \ldots, n \,.
\label{CartnBArelation}
\ee
This relation, which was proposed in \cite{Artz:1994qy},
is the same as for the closed $A^{(2)}_{2n}$ chain 
\cite{Reshetikhin:1987}. Its proof is sketched in Appendix \ref{sec:Cartan}.

The second part of the proof is the relation of the Dynkin label $[a_{1},
\ldots, a_{n}]$ to the  eigenvalues $(h_{1}, \ldots, h_{n})$ of the Cartan generators
\be
a_{i} &=& h_{i} - h_{i+1} \,, \qquad i = 1, 2, \ldots, n-1 \,, \non \\ 
a_{n} &=& \left\{ \begin{array}{rr}
2 h_{n} &  \mbox{ for } U_{q}(B_{n})\\
h_{n} &  \mbox{ for } U_{q}(C_{n})
\end{array}\right.
\,.
\label{DynkinCartnrelation}
\ee
This relation originates from the definition of Dynkin label (see 
e.g. \cite{Feger:2012bs})
\be
(h_{1}, \ldots, h_{n}) = \sum_{j=1}^{n} a_{j}\, \omega_{j} \,,
\label{Dynkindef}
\ee
where $\omega_{j}$ are the fundamental weights. In the orthogonal 
basis in which we work (recall Eqs. (\ref{simplerootsBn}),  
(\ref{simplerootsCn})), the 
fundamental weights are given by
\be
\omega_{1} &=& (1, 0 , 0,  0, \ldots, 0) \,, \non \\
\omega_{2} &=& (1, 1 , 0,  0, \ldots, 0) \,, \non \\
\omega_{3} &=& (1, 1 , 1,  0, \ldots, 0) \,, \non \\
&\vdots& \non \\
\omega_{n-1} &=& (1, 1 , 1, \ldots, 1, 0) \,, \non \\
\omega_{n} &=& \left\{ \begin{array}{rr}
(\frac{1}{2}, \frac{1}{2} , \ldots,  \frac{1}{2}) &  \mbox{ for } 
U_{q}(B_{n})\\
(1, 1 , \ldots,  1) &  \mbox{ for } U_{q}(C_{n})
\end{array}\right.
\,.
\ee 
Substituting these expressions for the fundamental weights into (\ref{Dynkindef}), we see that
\be
h_{1} &=& a_{1} + \ldots + a_{n-1} + \varepsilon\, a_{n} \,, \non \\
h_{2} &=& a_{2} + \ldots + a_{n-1} + \varepsilon\, a_{n} \,, \non \\
&\vdots& \non \\
h_{n} &=&  \varepsilon\, a_{n} \,, 
\label{harltns}
\ee
where 
\be
\varepsilon = \left\{ \begin{array}{rr}
\frac{1}{2} &  \mbox{ for } U_{q}(B_{n})\\
1 &  \mbox{ for } U_{q}(C_{n})
\end{array}\right.
\,.
\ee
Inverting the relations (\ref{harltns}), we arrive at the desired 
result (\ref{DynkinCartnrelation}).

The main result (\ref{DynkinBArltn}), (\ref{DynkinBArltn2})  follows immediately from the 
two relations (\ref{CartnBArelation}) and (\ref{DynkinCartnrelation}).

Since the Dynkin labels are nonnegative $a_{i} \ge 
0$, the result  (\ref{DynkinBArltn}) can be inverted to deduce the 
values of $(m_{1}, \ldots, m_{n})$ for which solutions of the Bethe 
equations (\ref{BAE}) with a given value of $N$ can be expected.

\section{Numerical check of completeness}\label{sec:completeness}

We present solutions ($\{u^{(1)}_{1}, \ldots, 
u^{(1)}_{m_{1}}\}, \ldots, \{u^{(n)}_{1}, \ldots, 
u^{(n)}_{m_{n}}\} $)
of the $A^{(2)}_{2n}$
Bethe equations (\ref{BAE}) for small 
values of $n$ and $N$ and a generic value of $\eta$ (namely, $\eta = 
-0.1 i$) in Tables \ref{table:B1N2} - \ref{table:B3N3} for set I  (\ref{KsetI}), 
and in Tables \ref{table:C1N2} - \ref{table:C3N3} for set II  (\ref{KsetII}). 
\footnote{The invariance of the Bethe equations under 
$u^{(l)}_{k} \mapsto u^{(l)}_{k} + 2\pi i$ and $u^{(l)}_{k} \mapsto 
-u^{(l)}_{k}$ can be used to restrict the Bethe 
roots to the domain $\Im m (u^{(l)}_{k}) \in [0, 2\pi)$ and $\Re e  
(u^{(l)}_{k}) \ge 0$.}
Each table also displays the cardinalities $(m_{1}, \ldots, m_{n})$ 
of the Bethe roots, 
the corresponding Dynkin label $[a_{1}, \ldots, a_{n}]$ obtained using the formula 
(\ref{DynkinBArltn}), the degeneracy (``deg'') of the corresponding 
eigenvalue of the Hamiltonians ${\cal H}^{(I)}$ and ${\cal H}^{(II)}$
(or, equivalently, of the transfer matrix $t(u)$ at some generic 
value of $u$) obtained by direct diagonalization, and the 
multiplicity (``mult'') i.e., the number of solutions of the Bethe equations 
with the given cardinality of Bethe roots.

We observe that, for each solution of the Bethe equations in these
tables, the dimension of the representation corresponding to the
Dynkin label coincides with the degeneracy. \footnote{The dimensions 
corresponding to the Dynkin labels can be read off from 
(\ref{decompB1})-(\ref{decompB3})  and 
(\ref{decompC1})-(\ref{decompC3}), or more generally can be obtained from e.g. 
\cite{Feger:2012bs}.}
Moreover, the degeneracies and multiplicities predicted by the quantum group 
symmetry (\ref{decompB1})-(\ref{decompB3}) and  
(\ref{decompC1})-(\ref{decompC3}) are completely accounted for by the 
Bethe ansatz solutions. \footnote{The astute reader will notice that 
two solutions are missing from 
Table \ref{table:C3N3}. We expect that this incompleteness can be 
attributed to our limited skill in finding solutions of nonlinear systems of 9 
equations with 9 unknowns, and not to the non-existence of such 
solutions.}

The eigenvalues of the Hamiltonians ${\cal H}^{(I)}$ 
(\ref{HamiltonianI}) and ${\cal H}^{(II)}$ 
(\ref{HamiltonianII}), as well as the eigenvalues of the transfer matrix $t(u)$ 
(\ref{transfer}) for the two sets (\ref{KsetI})-(\ref{KsetII}) at some generic value of $u$, are not displayed in 
the tables in order to minimize their size. Nevertheless, we have 
computed these eigenvalues both directly and from the reported 
solutions of the Bethe equations 
using (\ref{BAenergy}) and (\ref{Lambda})-(\ref{psi12}), 
respectively; and we find perfect agreement between the results from 
these two approaches.

\section{Conclusions}\label{sec:conclusion}

\begin{figure}
\centering
\subfloat[]{\includegraphics[width=5.5cm]{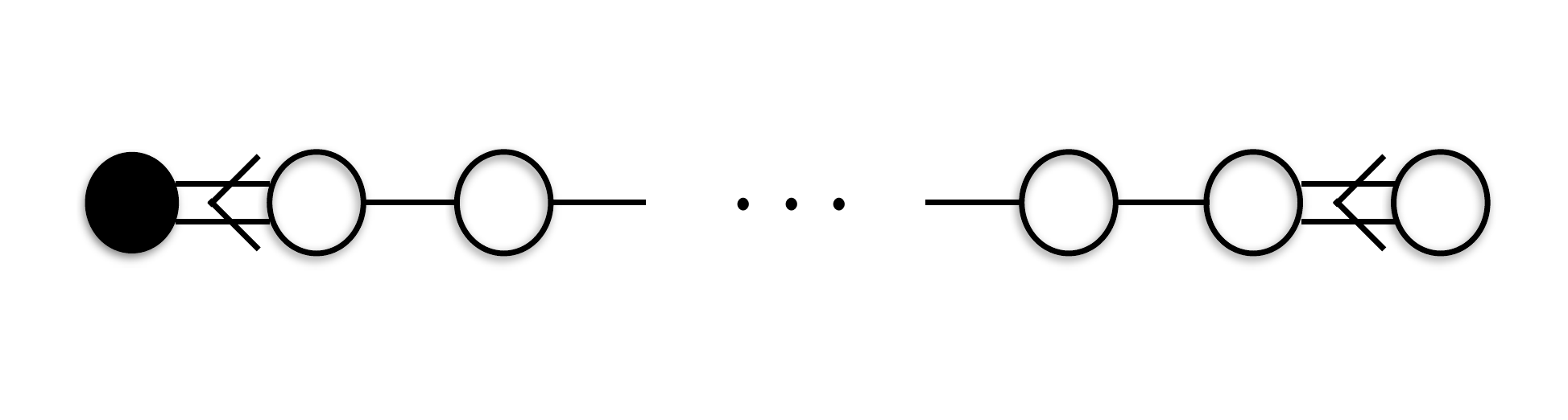}}
\subfloat[]{\includegraphics[width=5.5cm]{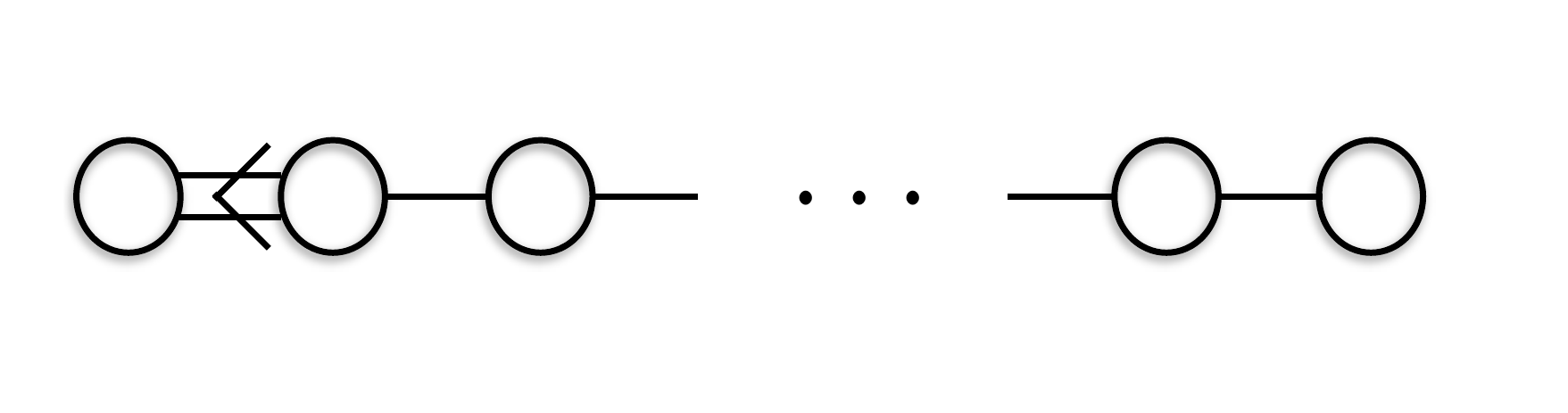}}
\subfloat[]{\includegraphics[width=5.5cm]{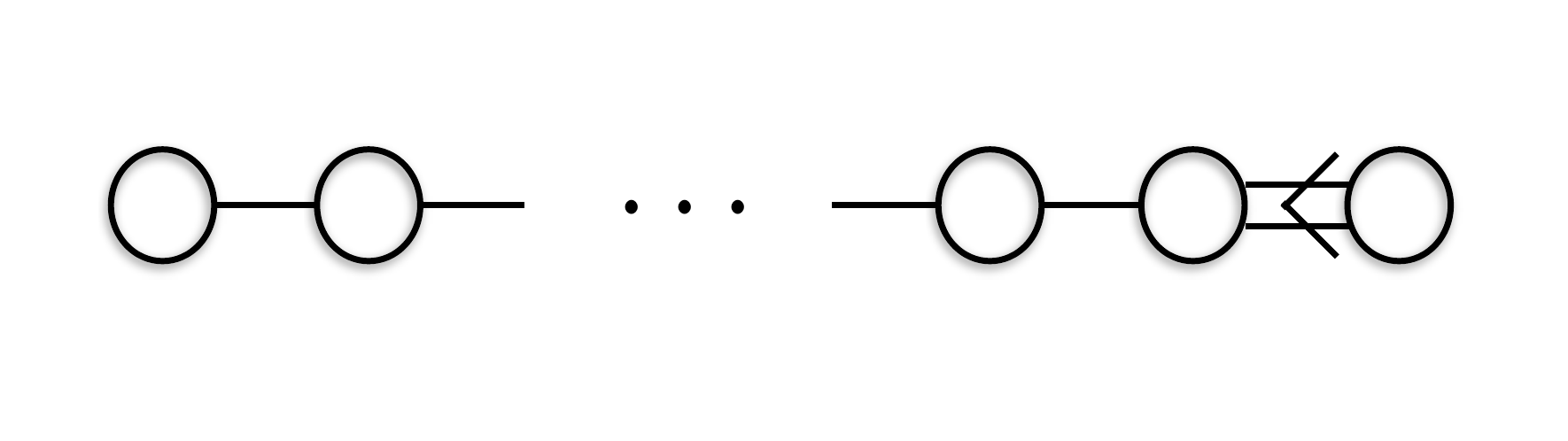}}
\caption{Dynkin diagrams for  
(a) $A^{(2)}_{2n}$ (b) $B_{n}$ (c) $C_{n}$}
\label{fig:Dynkin}
\end{figure}

We have argued that the $A_{2n}^{(2)}$ integrable open quantum spin
chains with the boundary conditions specified by (\ref{KsetI}) and (\ref{KsetII})
have the quantum group symmetries $U_{q}(B_{n})$ and $U_{q}(C_{n})$,
respectively, see Eqs. (\ref{UqBnsymmetry}) and  (\ref{UqCnsymmetry}).
A key point of this argument is that the Hamiltonians can be expressed as sums of 
two-body terms, see (\ref{HamiltonianI}) and (\ref{HamiltonianII}). 
In hindsight, the appearance of $B_{n}$ and $C_{n}$ can 
be inferred from the extended Dynkin diagram for $A^{(2)}_{2n}$ (see 
Fig. \ref{fig:Dynkin}): removing the rightmost or leftmost nodes 
yields the Dynkin diagrams for the subalgebras $B_{n}$ or $C_{n}$, respectively. 

We have also found a formula (\ref{DynkinBArltn}) for the
Dynkin label of a Bethe state; the Dynkin label uniquely
characterizes an irreducible representation, and in particular
determines its dimension, which is the degeneracy of the
corresponding eigenvalue. With the help of this formula,
we have verified numerically (for a generic value of $\eta$) that the 
degeneracies and multiplicities implied by the quantum group 
symmetry (\ref{decompB1})-(\ref{decompB3}) and  
(\ref{decompC1})-(\ref{decompC3}) are completely accounted for by the 
Bethe ansatz solutions, see Tables \ref{table:B1N2} - 
\ref{table:B3N3} and \ref{table:C1N2} - \ref{table:C3N3}, 
respectively. Similar results have recently been noted for the 
simpler case of the $U_{q}(A_{1})$-invariant spin-1/2 chain \cite{Pasquier:1989kd} 
at generic values of $q$ in \cite{Gainutdinov:2015vba}.

Several interesting problems remain to be addressed, including the 
following: 
proving that the transfer matrix $t(u)$ for the set II  (\ref{KsetII}) has 
$U_{q}(C_{n})$ symmetry; showing that the Bethe states have the 
highest weight property (\ref{highestweight}); and investigating the case 
that $q$ is a root of unity (non-generic values of $\eta$). We also 
note that the sets (\ref{KsetI}) and (\ref{KsetII}) do not exhaust 
the possible integrable diagonal boundary conditions 
\cite{Batchelor:1996np, LimaSantos:2002ui}. We expect that models with
these other boundary conditions will have ``less'' quantum group 
symmetry, which nevertheless may be worth exploring. It may also be 
interesting to find explicit formulas for the 
multiplicities in the tensor product decompositions of
$B_{n}$ (\ref{decomBn}) and $C_{n}$ (\ref{decomCn}) in terms of the Dynkin labels 
$[a_{1}, \ldots, a_{n}]$.\footnote{For the case of $A_{1}$, such a 
formula is well known, see e.g. Eq. (2.8) in 
\cite{Gainutdinov:2015vba}. Some recent progress on this problem was 
reported in \cite{Kulish2012a, Kulish2012b}.} 
These multiplicities should -- remarkably -- coincide with the number of solutions of 
the Bethe equations (\ref{BAE}) at generic values of $\eta$ for the corresponding  (\ref{DynkinBArltn})
values of $m_{1}, \ldots, m_{n}$.

\section*{Acknowledgments}
This paper is dedicated to the memory of Petr P. Kulish, who played an 
essential role in developing quantum groups and quantum 
integrability, the two main themes of this work.
One of us (RN) thanks Hubert Saleur for rekindling his interest in the 
symmetries of $A_{n}^{(2)}$ spin chains; Orlando Alvarez, Anastasia Doikou, Christian 
Korff, Rodrigo Pimenta, Kolya Reshetikhin, Luc Vinet, Robert Weston and Milen Yakimov for valuable discussions 
or correspondence; an anonymous referee for valuable comments; and
Robert Feger for pointing out \cite{Fonseca:2011sy}.
This work was supported in part by a Cooper fellowship.

\appendix

\section{The $A^{(2)}_{2n}$ R-matrix}

The R-matrix associated with the fundamental representation
of $A_{2n}^{(2)}$ was found by Bazhanov \cite{Bazhanov:1984gu, Bazhanov:1986mu} and
Jimbo \cite{Jimbo:1985ua}. We follow the latter reference; however,
as in \cite{Artz:1994qy}, we use the
variables $u$ and $\eta$ instead of $x$ and $k$, respectively, which
are related as follows:
\be
x = e^u \,, \qquad \qquad k = e^{2 \eta} \,.
\ee
The R-matrix is given by \footnote{This expression for the R-matrix
differs from the one given in Ref. \cite{Jimbo:1985ua} by the overall factor
$2 e^{u + (2n+3)\eta}$.}
\noindent
\be
R(u)&=&
c(u)\sum_{\alpha \neq \alpha'}e_{\alpha \alpha} \otimes e_{\alpha \alpha}
+ b(u) \sum_{\alpha \neq\beta,\beta'} e_{\alpha \alpha} \otimes
     e_{\beta \beta}  \non \\
&+& (e(u) \sum_{\alpha < \beta, \alpha \neq \beta'}
+ \bar e(u) \sum_{\alpha > \beta,\alpha\neq \beta'})\, e_{\alpha \beta}
\otimes e_{\beta \alpha}
+\sum_{\alpha \,, \beta} a_{\alpha \beta}(u)\, e_{\alpha \beta}
\otimes e_{\alpha' \beta'} \,,
\label{Rmatrix} 
\ee
with
\be
c(u)&=&2 \sinh(\uh-2\eta)\ \cosh(\uh - (2n+1)\eta) \,, \non \\
b(u)&=&2 \sinh(\uh)\ \cosh(\uh - (2n+1)\eta) \label{factors} \,, \\
e(u)&=&-2 e^{-\uh} \sinh (2\eta)\ \cosh(\uh -(2n+1)\eta) \non \,, \\
\bar e(u)&=&e^u e(u) \,, \non
\ee
\be
a_{\alpha \beta}(u)=\left\{ \begin{array}{ll}
\sinh(u-(2n-1)\eta)+\sinh((2n-1)\eta) & \alpha =\beta, \alpha \neq 
\alpha'\,, \\
\non\\
\sinh(u-(2n+1)\eta)+\sinh((2n+1)\eta)+\\+\sinh((2n-1)\eta)-\sinh((2n+3)\eta) &
\alpha=\beta, \alpha=\alpha' \,, \\
\non\\
-2 e^{((2n+1)+2(\bar{\alpha}-\bar{\beta}))\eta}e^{-\uh}\sinh\uh 
\sinh(2\eta) &
\alpha <\beta, \alpha\neq \beta' \,, \\
\non\\
2 e^{(2(2n+1)-2\beta+2)\eta} e^{-u}\sinh((2n+3-2 \beta)\eta) \sinh(2\eta)-\\-
2 e^{((2n+3)-2 \beta)\eta} \cosh((2(2n+2)-2 \beta)\eta) \sinh(2\eta) &
\alpha <\beta, \alpha=\beta' \,, \\
\non\\
2 e^{(-(2n+1)+2(\bar{\alpha}-\bar{\beta}))\eta} e^\uh \sinh \uh 
\sinh(2\eta) &
\alpha>\beta, \alpha\neq \beta' \,, \\
\non\\
2 e^{u-2 \beta\eta}\sinh (((2n+1)-2 \beta)\eta) \sinh(2\eta)-
\\-2 e^{((2n+1)-2\beta)\eta} 
\cosh(2\beta\eta) \sinh(2\eta) & \alpha>\beta, \alpha=\beta' \,,
\end{array} \right. 
\ee
where
\be
\bar{\alpha}=\left\{ \begin{array}{ll}
\alpha+\half & 1\le\alpha<n+1\\
\alpha & \alpha=n+1\\
\alpha-\half & n+1<\alpha\le 2n+1
\end{array} \right. \,,
\ee
\be
\alpha' &=& 2n+2-\alpha  \,, \non \\
\alpha,\beta &=& 1 \,, 2\,, \ldots \,, 2n+1 \,.
\ee

This R-matrix has crossing symmetry (\ref{crossing}), where $V$ is 
given by \footnote{We take this opportunity to correct several typos in 
the corresponding equation (59) in \cite{Artz:1994qy}.}
\be
V = \sum_{\alpha} e_{\alpha 
\alpha}\delta_{\alpha,\alpha'}+\sum_{\alpha<\alpha'}e^{[-(2n+1)+2\alpha]\eta}e_{\alpha \alpha'}
+\sum_{\alpha>\alpha'}e^{(2n+1-2\alpha')\eta}e_{\alpha \alpha'} \,.
\label{Vmat}
\ee
The matrix $M = V^{t} V$ is therefore given by the diagonal matrix
\be
M  = \diag(e^{4(n+1-\bar\alpha)\eta})\,, \qquad \alpha = 1, 2, \ldots \,, 2n+1 \,.
\label{Mmat}
\ee

\section{Eigenvalues of the Cartan generators}\label{sec:Cartan}

We sketch here a proof of the relation (\ref{CartnBArelation})
\be
h_{1} &=& N - m_{1} \,, \non \\
h_{i} &=& m_{i-1} - m_{i} \,, \qquad i = 2, 3, \ldots, n \,,
\label{CartnBArelation2}
\ee
based on the nested algebraic Bethe ansatz solution \cite{Li:2005pp}.
Since the argument is somewhat intricate, it is helpful to first consider some special
cases.  Hence, as a first warm-up, we consider the case $A^{(2)}_{2}$
in Section \ref{sec:warmup1}; and then, as a second warm-up, we
consider the case $A^{(2)}_{4}$ in Section \ref{sec:warmup2}.
Finally, we consider the general case $A^{(2)}_{2n}$ in Section
\ref{sec:general}. \footnote{The proof of (\ref{CartnBArelation2}) presented here supersedes 
the discussion given in Appendix B of \cite{Artz:1994qy}.}

\subsection{$A^{(2)}_{2}$}\label{sec:warmup1}

For the case $n=1$, the Bethe states are given by
\be
|\Lambda^{(m_{1})} \rangle = B_{2}(u^{(1)}_{1}) \cdots 
B_{2}(u^{(1)}_{m_{1}}) |0\rangle + \ldots \,,
\label{A2level1}
\ee
where $B_{2}(u)$ is the operator appearing in the double-row 
monodromy matrix (\ref{doublerowmonodromy}), and $|0\rangle$ is the 
reference state (\ref{reference}). The ellipsis denotes contributions 
from terms that also depend on the operator $F(u)$, which here and 
below we assume can be safely ignored.
Using the facts \footnote{In order to lighten the notation,
here and below we drop the notation $\Delta_{(k)}$ for the Cartan 
generators on $k$ sites.}
\be
\left[ H_{1}\,, B_{2}(u) \right] = - B_{2}(u)\,, \qquad  H_{1} 
|0\rangle = N |0\rangle \,,
\ee
we immediately see that
\be
H_{1} |\Lambda^{(m_{1})} \rangle = (N-m_{1}) |\Lambda^{(m_{1})} 
\rangle \,.
\label{A2result}
\ee
Therefore $h_{1} = N-m_{1}$, in agreement with (\ref{CartnBArelation2}).

\subsection{$A^{(2)}_{4}$}\label{sec:warmup2}

We now consider the case $n=2$, where nesting first appears. The 
(first-level) Bethe states are 
given by
\be
|\Lambda^{(m_{1}, m_{2})} \rangle = f_{i_{1} \cdots i_{m_{1}}} B_{i_{1}}(u^{(1)}_{1}) \cdots 
B_{i_{m_{1}}}(u^{(1)}_{m_{1}}) |0\rangle + \ldots \,,
\label{A4level1}
\ee
where $i_{1}, \ldots,  i_{m_{1}} \in \{2, 3, 4\}$, 
$f_{i_{1} \cdots i_{m_{1}}}$ are coefficients that are still to be 
determined, and summation over repeated indices is understood.

Let $n_{i}$ denote the number of $B_{i}(u)$ operators appearing in 
$|\Lambda^{(m_{1}, m_{2})} \rangle$ (\ref{A4level1}). Evidently,
\be
m_{1} = n_{2} + n_{3} + n_{4} \,.
\label{evident}
\ee
Using the facts
\be
\left[ H_{1}\,, B_{j}(u) \right] = - B_{j}(u)\,, \qquad  j = 2, 3, 
4\,, \qquad H_{1} |0\rangle = N |0\rangle \,,
\ee
we obtain
\be
H_{1} |\Lambda^{(m_{1}, m_{2})} \rangle = 
(N-n_{2}-n_{3}-n_{4}) |\Lambda^{(m_{1}, m_{2})} \rangle\,,
\ee
which, in view of (\ref{evident}), again implies $h_{1} = N-m_{1}$.

Moreover, using the facts
\be
\left[ H_{2}\,, B_{j}(u) \right] = \left\{\begin{array}{rr}
B_{j}(u) &  \mbox{ for }  j=2\\
- B_{j}(u)&  \mbox{ for } j=4\\
0 & \mbox{ otherwise }
\end{array}\right.
\,, \qquad  H_{2} |0\rangle = 0  \,,
\ee
we obtain
\be
H_{2} |\Lambda^{(m_{1}, m_{2})} \rangle = 
(n_{2}-n_{4}) |\Lambda^{(m_{1}, m_{2})} \rangle \,,
\ee
which implies 
\be
h_{2} = n_{2}-n_{4}\,.
\label{h2A4}
\ee

The coefficients in (\ref{A4level1}) are given by the scalar product 
\footnote{Since the transfer matrix is symmetric (see Appendix B in 
\cite{Mezincescu:1991ag}), its left and right eigenvectors are each other's transpose.}
\be
f_{i_{1} \cdots i_{m_{1}}} = \left( \langle e_{i_{1}}| \otimes \cdots \otimes
\langle e_{i_{m_{1}}}| \right) |\tilde \psi \rangle \,,
\ee
where $|\tilde \psi \rangle $ is the second-level state
\be
|\tilde \psi \rangle = \tilde B_{2}(u^{(2)}_{1}) \cdots 
\tilde B_{2}(u^{(2)}_{m_{2}}) |\tilde 0\rangle + \ldots 
\ee
where $\tilde B_{2}(u)$ are the $A^{(2)}_{2}$ creation operators 
constructed as in (\ref{scriptT}) with $n=1$ except
with {\em inhomogeneous} monodromy matrices (the inhomogeneities are 
given by $\{ u^{(1)}_{1}\,, \ldots \,, u^{(1)}_{m_{1}} \}$). 
Moreover,
\be
|\tilde 0\rangle =\left(\begin{array}{c}
1 \\
0 \\
0
\end{array}\right)^{\otimes m_{1}}\,, 
\ee
and
\be
|e_{2} \rangle=  \left(\begin{array}{c}
1 \\
0 \\
0
\end{array}\right) \,, \qquad 
|e_{3} \rangle=  \left(\begin{array}{c}
0 \\
1 \\
0
\end{array}\right) \,, \qquad 
|e_{4}\rangle =  \left(\begin{array}{c}
0 \\
0 \\
1
\end{array}\right) \,.
\ee

Let $\tilde H_{1}$ denote the Cartan generator for the case  
$A^{(2)}_{2}$, and let us now evaluate its matrix element
\be
\left( \langle e_{i_{1}}| \otimes \cdots \otimes
\langle e_{i_{m_{1}}}| \right) \tilde H_{1} |\tilde \psi \rangle 
\ee
in two different ways. To compute the action of $\tilde H_{1} $ to the right, we 
use $\tilde H_{1} |\tilde \psi \rangle  = (m_{1} - m_{2})  |\tilde 
\psi \rangle$, similarly to (\ref{A2result}). To compute the action 
of  $\tilde H_{1}$ to the left, we use the fact
\be
\tilde H_{1}  |e_{j}\rangle = \left\{\begin{array}{rr}
|e_{j}\rangle &  \mbox{ for }  j=2\\
- |e_{j}\rangle &  \mbox{ for } j=4\\
0 & \mbox{ otherwise }
\end{array}\right.
\,,
\ee
and therefore
\be
\left( \langle e_{i_{1}}| \otimes \cdots \otimes
\langle e_{i_{m_{1}}}| \right) \tilde H_{1} = \left( \langle e_{i_{1}}| \otimes \cdots \otimes
\langle e_{i_{m_{1}}}| \right)  (n_{2} - n_{4}) \,.
\ee
We conclude that
\be
(n_{2} - n_{4})\, f_{i_{1} \cdots i_{m_{1}}} =    (m_{1} - m_{2})\, 
f_{i_{1} \cdots i_{m_{1}}} \,,
\ee
which implies that $f_{i_{1} \cdots i_{m_{1}}}$ is zero unless
\be
n_{2} - n_{4} = m_{1} - m_{2} \,.
\ee
Recalling (\ref{h2A4}), we conclude that $h_{2} = m_{1} - m_{2} $, in 
agreement with (\ref{CartnBArelation2}).

\subsection{$A^{(2)}_{2n}$}\label{sec:general}

In order to treat the general case, it is necessary to adopt a more 
systematic (but unfortunately significantly heavier) notation. 
We therefore write the Bethe states as
\be
|\Lambda^{(m_{1}, \ldots, m_{n})} \rangle = f^{(1)}_{i^{(1)}_{1} 
\cdots i^{(1)}_{m_{1}}} |\psi^{(1)}\rangle_{i^{(1)}_{1} 
\cdots i^{(1)}_{m_{1}}} \,.
\label{A2nBethe}
\ee

\subsubsection{First level}

The first-level states are given by
\be
|\psi^{(1)}\rangle_{i^{(1)}_{1} \cdots i^{(1)}_{m_{1}}}  = 
B^{(1)}_{i^{(1)}_{1}}(u^{(1)}_{1}) \cdots 
B^{(1)}_{i^{(1)}_{m_{1}}}(u^{(1)}_{m_{1}}) |0^{(1)}\rangle + \ldots \,,
\label{A2nlevel1}
\ee
where $i^{(1)}_{1}, \ldots,  i^{(1)}_{m_{1}} \in \{2, \ldots, 2n\}$;
and $B^{(1)}_{i}(u) \equiv B_{i}(u)$ and $|0^{(1)}\rangle \equiv 
|0 \rangle$ are given by (\ref{doublerowmonodromy}) and  (\ref{reference}), respectively.

Letting $n^{(1)}_{i}$ denote the number of $B^{(1)}_{i}(u)$ operators 
appearing in (\ref{A2nlevel1}), we have
\be
m_{1} = n^{(1)}_{2} + \ldots + n^{(1)}_{2n} \,.
\ee
For $H^{(1)}_{i} \equiv H_{i}$, we have for $i=1$:
\be
\left[ H^{(1)}_{1}\,, B^{(1)}_{j}(u) \right] = - B^{(1)}_{j}(u)\,, \qquad  
j = 2, \ldots, 2n \,, \qquad H^{(1)}_{1} |0^{(1)}\rangle = N 
|0^{(1)}\rangle \,;
\ee
and for $i>1$:
\be
\left[ H^{(1)}_{i}\,, B^{(1)}_{j}(u) \right] = \left\{\begin{array}{cl}
B^{(1)}_{j}(u)  &  \mbox{ for }  j=i\\
- B^{(1)}_{j}(u) &  \mbox{ for } j=2n+2-i\\
0 & \mbox{ otherwise }
\end{array}\right.
\,, \qquad  H^{(1)}_{i} |0^{(1)}\rangle = 0  \,.
\ee
Therefore
\be
H^{(1)}_{1} |\Lambda^{(m_{1},\ldots, m_{n})} \rangle &=& 
(N-n^{(1)}_{2}-\ldots-n^{(1)}_{2n})\,  |\Lambda^{(m_{1},\ldots, m_{n})} 
\rangle \,, \non \\
H^{(1)}_{i} |\Lambda^{(m_{1},\ldots, m_{n})} \rangle &=& 
(n^{(1)}_{i}-n^{(1)}_{2n+2-i})\,  |\Lambda^{(m_{1},\ldots, 
m_{n})} \rangle \,, \qquad i = 2, \ldots, n \,,
\ee
which implies 
\be
h_{1} &=& N-m_{1} \,, \non \\
h_{i} &=& n^{(1)}_{i}-n^{(1)}_{2n+2-i} \,, \qquad i = 2, 
\ldots, n \,.
\label{resultlev1}
\ee

\subsubsection{Second level}

The coefficients in (\ref{A2nBethe}) are given by the scalar product
\be
f^{(1)}_{i^{(1)}_{1} 
\cdots i^{(1)}_{m_{1}}} = \left( \langle e^{(1)}_{i^{(1)}_{1}}| \otimes \cdots \otimes
\langle e^{(1)}_{i^{(1)}_{m_{1}}}| \right) |\psi^{(2)}\rangle_{i^{(2)}_{1} 
\cdots i^{(2)}_{m_{2}}}\,  f^{(2)}_{i^{(2)}_{1} \cdots i^{(2)}_{m_{2}}}
\,,
\ee
where the second-level states are given by
\be
|\psi^{(2)}\rangle_{i^{(2)}_{1} \cdots i^{(2)}_{m_{2}}}  = 
B^{(2)}_{i^{(2)}_{1}}(u^{(2)}_{1}) \cdots 
B^{(2)}_{i^{(2)}_{m_{2}}}(u^{(2)}_{m_{2}}) |0^{(2)}\rangle + \ldots \,,
\label{A2nlevel2}
\ee
where $i^{(2)}_{1}, \ldots,  i^{(2)}_{m_{2}} \in \{2, \ldots, 2n-2\}$;
$B^{(2)}_{i}(u)$ are the (inhomogeneous) creation operators for 
$A^{(2)}_{2n-2}$; and
\be
|0^{(2)}\rangle =\left(\begin{array}{c}
1 \\
0 \\
\vdots \\
0
\end{array}\right)_{2n-1}^{\otimes m_{1}}\,.
\ee
Moreover,
\be
|e^{(1)}_{2} \rangle=  \left(\begin{array}{c}
1 \\
0 \\
\vdots \\
0
\end{array}\right)_{2n-1} \,, \qquad \ldots  \qquad \,, \qquad
|e^{(1)}_{2n}\rangle =  \left(\begin{array}{c}
0 \\
\vdots \\
0 \\
1
\end{array}\right)_{2n-1} \,.
\ee

We have that 
\be
m_{2} = n^{(2)}_{2} + \ldots + n^{(2)}_{2n-2} \,,
\ee
and hence
\be
H^{(2)}_{1} |\psi^{(2)}\rangle_{i^{(2)}_{1} \cdots i^{(2)}_{m_{2}}}  &=& 
(m_{1} - m_{2})\,  |\psi^{(2)}\rangle_{i^{(2)}_{1} \cdots i^{(2)}_{m_{2}}}\,, \non \\
H^{(2)}_{i} |\psi^{(2)}\rangle_{i^{(2)}_{1} \cdots i^{(2)}_{m_{2}}} &=& 
(n^{(2)}_{i}-n^{(2)}_{2n-i})\,  |\psi^{(2)}\rangle_{i^{(2)}_{1} 
\cdots i^{(2)}_{m_{2}}} \,, \qquad i = 2, \ldots, n-1 \,.
\ee
Furthermore,
\be
H^{(2)}_{i}  |e^{(1)}_{j}\rangle = \left\{\begin{array}{rl}
|e^{(1)}_{j}\rangle &  \mbox{ for }  j=i+1\\
- |e^{(1)}_{j}\rangle &  \mbox{ for } j=2n+1-i\\
0 & \mbox{ otherwise }
\end{array}\right.
\,.
\ee
Evaluating the matrix element
\be
\left( \langle e^{(1)}_{i^{(1)}_{1}}| \otimes \cdots \otimes
\langle e^{(1)}_{i^{(1)}_{m_{1}}}| \right)  H^{(2)}_{i}
|\psi^{(2)}\rangle_{i^{(2)}_{1} 
\cdots i^{(2)}_{m_{2}}}\,  f^{(2)}_{i^{(2)}_{1} \cdots 
i^{(2)}_{m_{2}}}
\label{matelem}
\ee
in two different ways by acting with $ H^{(2)}_{i}$ to both the left 
and the right, we obtain for $i=1$
\be
n^{(1)}_{2} - n^{(1)}_{2n} = m_{1} - m_{2} \,,
\label{resultlev2a}
\ee
and for $i>1$
\be
n^{(1)}_{i+1} - n^{(1)}_{2n+1-i} = n^{(2)}_{i} - n^{(2)}_{2n-i} \,, 
\qquad i = 2, \ldots, n-1 \,.
\label{resultlev2b}
\ee

\subsubsection{Level $k$}

At level $k = 2, 3, \ldots, n-1$, we have
\be
f^{(k-1)}_{i^{(k-1)}_{1} 
\cdots i^{(k-1)}_{m_{k-1}}} = \left( \langle e^{(k-1)}_{i^{(k-1)}_{1}}| \otimes \cdots \otimes
\langle e^{(k-1)}_{i^{(k-1)}_{m_{k-1}}}| \right) 
|\psi^{(k)}\rangle_{i^{(k)}_{1} 
\cdots i^{(k)}_{m_{k}}}\,  f^{(k)}_{i^{(k)}_{1} \cdots i^{(k)}_{m_{k}}}
\,,
\ee
where the level-$k$ states are given by
\be
|\psi^{(k)}\rangle_{i^{(k)}_{1} \cdots i^{(k)}_{m_{k}}}  = 
B^{(k)}_{i^{(k)}_{1}}(u^{(k)}_{1}) \cdots 
B^{(k)}_{i^{(k)}_{m_{k}}}(u^{(k)}_{m_{k}}) |0^{(k)}\rangle + \ldots \,,
\label{A2nlevelk}
\ee
where $i^{(k)}_{1}, \ldots,  i^{(k)}_{m_{k}} \in \{2, \ldots, 2n-2k+2\}$;
$B^{(k)}_{i}(u)$ are the (inhomogeneous) creation operators for 
$A^{(2)}_{2n-2k+2}$; and
\be
|0^{(k)}\rangle =\left(\begin{array}{c}
1 \\
0 \\
\vdots \\
0
\end{array}\right)_{2n-2k+3}^{\otimes m_{k-1}}\,.
\ee
Moreover,
\be
|e^{(k-1)}_{2} \rangle=  \left(\begin{array}{c}
1 \\
0 \\
\vdots \\
0
\end{array}\right)_{2n-2k+3} \,, \qquad \ldots  \qquad \,, \qquad
|e^{(k-1)}_{2n-2k+4}\rangle =  \left(\begin{array}{c}
0 \\
\vdots \\
0 \\
1
\end{array}\right)_{2n-2k+3} \,.
\ee

We have that 
\be
m_{k} = n^{(k)}_{2} + \ldots + n^{(k)}_{2n-2k+2} \,.
\ee
Also,
\be
\left[ H^{(k)}_{1}\,, B^{(k)}_{j}(u) \right] = - B^{(k)}_{j}(u)\,, \quad  
j = 2, \ldots, 2n-2k+2 \,, \quad H^{(k)}_{1} |0^{(k)}\rangle = m_{k-1} 
|0^{(k)}\rangle \,;
\ee
and for $i>1$:
\be
\left[ H^{(k)}_{i}\,, B^{(k)}_{j}(u) \right] = \left\{\begin{array}{cl}
B^{(k)}_{j}(u)  &  \mbox{ for }  j=i\\
- B^{(k)}_{j}(u) &  \mbox{ for } j=2n-2k+4-i\\
0 & \mbox{ otherwise }
\end{array}\right.
\,, \qquad  H^{(k)}_{i} |0^{(k)}\rangle = 0  \,.
\ee
Hence
\be
H^{(k)}_{1} |\psi^{(k)}\rangle_{i^{(k)}_{1} \cdots i^{(k)}_{m_{k}}}  &=& 
(m_{k-1} - m_{k})\,  |\psi^{(k)}\rangle_{i^{(k)}_{1} \cdots 
i^{(k)}_{m_{k}}}\,, \\
H^{(k)}_{i} |\psi^{(k)}\rangle_{i^{(k)}_{1} \cdots i^{(k)}_{m_{k}}} &=& 
(n^{(k)}_{i}-n^{(k)}_{2n-2k+4-i})\,  |\psi^{(k)}\rangle_{i^{(k)}_{1} 
\cdots i^{(k)}_{m_{k}}} \,, \qquad i = 2, \ldots, n-k+1 \,.  \non
\ee
Furthermore,
\be
H^{(k)}_{i}  |e^{(k-1)}_{j}\rangle = \left\{\begin{array}{rl}
|e^{(k-1)}_{j}\rangle &  \mbox{ for }  j=i+1\\
- |e^{(k-1)}_{j}\rangle &  \mbox{ for } j=2n-2k+5-i\\
0 & \mbox{ otherwise }
\end{array}\right.
\,.
\ee
Evaluating the matrix element
\be
\left( \langle e^{(k-1)}_{i^{(k-1)}_{1}}| \otimes \cdots \otimes
\langle e^{(k-1)}_{i^{(k-1)}_{m_{k-1}}}| \right)  H^{(k)}_{i}
|\psi^{(k)}\rangle_{i^{(k)}_{1} 
\cdots i^{(k)}_{m_{k}}}\,  f^{(k)}_{i^{(k)}_{1} \cdots 
i^{(k)}_{m_{k}}} 
\ee
in two different ways by acting with $ H^{(k)}_{i}$ to both the left 
and the right, we obtain 
\be
n^{(k-1)}_{2} - n^{(k-1)}_{2n-2k+4} &=& m_{k-1} - m_{k} \,, \non \\
n^{(k-1)}_{i+1} - n^{(k-1)}_{2n-2k+5-i} &=& n^{(k)}_{i} - 
n^{(k)}_{2n-2k+4-i} \,, 
\qquad i = 2, \ldots, n-k+1 \,.
\label{resultlevk}
\ee

\subsubsection{Level $n$}

At the final level $k=n$, we have
\be
f^{(n-1)}_{i^{(n-1)}_{1} 
\cdots i^{(n-1)}_{m_{n-1}}} = \left( \langle e^{(n-1)}_{i^{(n-1)}_{1}}| \otimes \cdots \otimes
\langle e^{(n-1)}_{i^{(n-1)}_{m_{n-1}}}| \right) 
|\psi^{(n)}\rangle
\,,
\ee
where the level-$n$ states are given by
\be
|\psi^{(n)}\rangle  = 
B^{(n)}_{2}(u^{(n)}_{1}) \cdots 
B^{(n)}_{2}(u^{(n)}_{m_{n}}) |0^{(n)}\rangle + \ldots \,,
\label{A2nleveln}
\ee
where $B^{(n)}_{i}(u)$ are the (inhomogeneous) creation operators for 
$A^{(2)}_{2}$. Moreover,
\be
|0^{(n)}\rangle =\left(\begin{array}{c}
1 \\
0 \\
0
\end{array}\right)^{\otimes m_{n-1}}\,,
\ee
and
\be
|e^{(n-1)}_{2} \rangle=  \left(\begin{array}{c}
1 \\
0 \\
0
\end{array}\right) \,, \qquad 
|e^{(n)}_{3} \rangle=  \left(\begin{array}{c}
0 \\
1 \\
0
\end{array}\right) \,, \qquad 
|e^{(n)}_{4}\rangle =  \left(\begin{array}{c}
0 \\
0 \\
1
\end{array}\right) \,.
\ee

We have that 
\be
H^{(n)}_{1} |\psi^{(n)}\rangle  = (m_{n-1} - m_{n})\,  |\psi^{(n)}\rangle
\ee
and
\be
H^{(n)}_{1}  |e^{(n-1)}_{j}\rangle = \left\{\begin{array}{rl}
|e^{(n-1)}_{j}\rangle &  \mbox{ for }  j=2\\
- |e^{(n-1)}_{j}\rangle &  \mbox{ for } j=4\\
0 & \mbox{ otherwise }
\end{array}\right.
\,.
\ee
Evaluating the matrix element
\be
\left( \langle e^{(n-1)}_{i^{(n-1)}_{1}}| \otimes \cdots \otimes
\langle e^{(n-1)}_{i^{(n-1)}_{m_{n-1}}}| \right)  H^{(n)}_{1}
|\psi^{(n)}\rangle
\ee
in two different ways, we obtain 
\be
n^{(n-1)}_{2} - n^{(n-1)}_{4} = m_{n-1} - m_{n} \,.
\label{resultlevn}
\ee

Combining all the results (\ref{resultlev1}), (\ref{resultlev2a}), 
(\ref{resultlev2b}), (\ref{resultlevk}), (\ref{resultlevn}), we 
obtain the desired relations (\ref{CartnBArelation2}). Indeed,
one can see that
\be
h_{i} = n^{(k-1)}_{i+2-k} - n^{(k-1)}_{2n+4-k-i}\,, \qquad k=2, \ldots, 
n\,,
\ee
which gives for $i=k$
\be
h_{k} = n^{(k-1)}_{2} - n^{(k-1)}_{2n-2k+4} = m_{k-1} - m_{k}\,,
\ee
where the second equality follows from  (\ref{resultlevk}).

\newpage
\clearpage

\begin{table}
\centering
\begin{tabular}{|c|c|c|c|c|}
\hline
$m_{1}$ & $a_{1}$ & deg & mult & $\{ u^{(1)}_{k} \}$\\   
\hline
0 & 4  & 5 & 1 & - \\
\hline
1 & 2 & 3 & 1 & 0.201347 \\
\hline
2 & 0  & 1 & 1 & $0.627218 \pm 1.28621 i$\\
\hline
\end{tabular}
\caption{\small $U_{q}(B_{1})$, $N=2$}\label{table:B1N2}
\end{table}

\begin{table}
\centering
\begin{tabular}{|c|c|c|c|c|}
\hline
$m_{1}$ & $a_{1}$ & deg & mult & $\{ u^{(1)}_{k} \}$\\   
\hline
0 & 6  & 7 & 1 & - \\
\hline
1 & 4  & 5 & 2 & 0.115986 \\
  &    &   &   & 0.351133 \\
\hline
2 & 2  & 3 & 3 & $0.524753 \pm 1.38161 i$ \\
  &    &   &  & 0.11483\,, $1.56044 i$ \\
  &    &   &  & 0.343261\,, $1.64011 i$ \\
\hline
3 & 0  & 1 & 1 & 0.115223\,, 0.344343\,, 0.324313 +$i \pi$  \\
\hline 
\end{tabular}
\caption{\small $U_{q}(B_{1})$, $N=3$}\label{table:B1N3}
\end{table}

\begin{table}
\centering
\begin{tabular}{|c|c||c|c||c|c|c|c|}
\hline
$m_{1}$ & $m_{2}$ & $a_{1}$ & $a_{2}$ & deg & mult & $\{ u^{(1)}_{k} 
\}$ & $\{ u^{(2)}_{k} \}$\\   
\hline
0 & 0 & 2 & 0  & 14 & 1 & - & - \\
\hline
1 & 0 & 0 & 2  & 10 & 1 & 0.201347  & - \\
\hline
2 & 2 & 0 & 0  & 1 & 1 & $0.427307 \pm 0.971435 i$  &  $0.506682 \pm 
1.38565 i$\\
\hline
\end{tabular}
\caption{\small $U_{q}(B_{2})$, $N=2$}\label{table:B2N2}
\end{table}

\begin{table}
\centering
\begin{tabular}{|c|c||c|c||c|c|c|c|}
\hline
$m_{1}$ & $m_{2}$ & $a_{1}$ & $a_{2}$ & deg & mult & $\{ u^{(1)}_{k} 
\}$ & $\{ u^{(2)}_{k} \}$\\   
\hline
0 & 0 & 3 & 0  & 30 & 1 & - & - \\
\hline
1 & 0 & 1 & 2  & 35 & 2 & 0.115986  & - \\
 &  &  &   &  &  & 0.351133  & - \\
\hline
2 & 1 & 0 & 2  & 10 & 1 & 0.115986\,, 0.351133  & 0.331791 \\
\hline
2 & 2 & 1 & 0  & 5 & 3 & 0.338012\,, $1.11733 i$  & $0.340113 \pm 
1.32976 i$ \\
 &  &  &   &  &  & $0.390693 \pm 1.11745 i$ & $0.434061 \pm 1.4248 i$ \\
 &  &  &   &  &  &  0.113154\,, $1.01242 i$ & $0.410526 \pm 1.3294 i$ \\
\hline
\end{tabular}
\caption{\small $U_{q}(B_{2})$, $N=3$}\label{table:B2N3}
\end{table}

\begin{table}
\small 
\centering
\begin{tabular}{|c|c|c||c|c|c||c|c|c|c|c|}
\hline
$m_{1}$ & $m_{2}$ & $m_{3}$ & $a_{1}$ & $a_{2}$  & $a_{3}$ & deg & mult & $\{ u^{(1)}_{k} 
\}$ & $\{ u^{(2)}_{k} \}$ & $\{ u^{(3)}_{k} \}$\\   
\hline
0 & 0 & 0 & 2 & 0 & 0  & 27 & 1 & - & - & - \\
\hline
1 & 0 & 0 & 0 & 1 & 0  & 21 & 1 & 0.201347  & - & - \\
\hline
2 & 2 & 2 & 0 & 0 & 0  & 1 & 1 & $0.327207 \pm 0.786874 i$
& $0.372666 \pm 1.12783 i$  &  $0.415697 \pm 1.42783 i$\\
\hline
\end{tabular}
\caption{\small $U_{q}(B_{3})$, $N=2$}\label{table:B3N2}
\end{table}

\begin{table}
\small
\centering
\begin{tabular}{|c|c|c||c|c|c||c|c|c|c|c|}
\hline
$m_{1}$ & $m_{2}$ & $m_{3}$ & $a_{1}$ & $a_{2}$  & $a_{3}$ & deg & mult & $\{ u^{(1)}_{k} 
\}$ & $\{ u^{(2)}_{k} \}$ & $\{ u^{(3)}_{k} \}$\\   
\hline
0 & 0 & 0 & 3 & 0 & 0  & 77 & 1 & - & - & - \\
\hline
1 & 0 & 0 & 1 & 1 & 0  & 105 & 2 & 0.115986  & - & - \\
 &  &  &  &  &   &  &  & 0.351133  & - & - \\
\hline
2 & 1 & 0 & 0 & 0 & 2  & 35 & 1 & 0.115986\,, 0.351133  & 0.331791 & - \\
\hline
2 & 2 & 2 & 1 & 0 & 0  & 7 & 3 & 0.110446\,, $0.776613 i$  & 
$0.287874 \pm 1.03712 i$ & $0.387205 \pm 1.40363 i$ \\
 &  &  &  &  &   &  &  & 0.335123\,, $0.905482 i$  & $0.186397 \pm 
 1.03899 i$ & $0.327378 \pm 1.40468 i$ \\
 &  &  &  &  &   &  &  & $0.308473 \pm 0.927961 i$  & $0.333096 \pm 
 1.19533 i$ & $0.358468 \pm 1.44765 i$ \\
\hline
\end{tabular}
\caption{\small $U_{q}(B_{3})$, $N=3$}\label{table:B3N3}
\end{table}

\begin{table}
\centering
\begin{tabular}{|c|c|c|c|c|}
\hline
$m_{1}$ & $a_{1}$ & deg & mult & $\{ u^{(1)}_{k} \}$\\   
\hline
0 & 2  & 3 & 1 & - \\
\hline
1 & 1 & 2 & 2 & 0.185137 \\
 &  &  &  & 1.04997 \\
\hline
2 & 0  & 1 & 2 & $0.757565  \pm 0.363991  i$\\
 &   &  &  &  0.206122\,, $2.59788 i$\\
\hline
\end{tabular}
\caption{\small $U_{q}(C_{1})$, $N=2$}\label{table:C1N2}
\end{table}

\begin{table}
\centering
\begin{tabular}{|c|c|c|c|c|}
\hline
$m_{1}$ & $a_{1}$ & deg & mult & $\{ u^{(1)}_{k} \}$\\   
\hline
0 & 3  & 4 & 1 & - \\
\hline
1 & 2  & 3 & 3 & 0.111524 \\
  &    &   &   & 0.315352 \\
    &    &   &   & 1.38581 \\
\hline
2 & 1  & 2 & 5 & $1.10381  \pm 0.414939  i$ \\
  &    &   &  & 0.116934\,, 0.776633 \\
  &    &   &  & 0.454616\,, 0.531061 \\
  &    &   &  & 0.117801\,, $2.59116 i$ \\
  &    &   &  & 0.369036\,, $2.73713 i$ \\
\hline
3 & 0  & 1 & 4 & 0.886562\,, $0.777865\pm 0.638435  i$ \\
 &   &   &   & 0.417895\,, 0.773051\,, $2.5569 i$ \\
 &   &   &   & 0.119916\,, 0.88464\,, $2.47305 i$ \\
 &   &   &   & 0.113539\,, 0.333831\,, 0.365549 + $i \pi $ \\
\hline 
\end{tabular}
\caption{\small $U_{q}(C_{1})$, $N=3$}\label{table:C1N3}
\end{table}

\begin{table}
\centering
\begin{tabular}{|c|c||c|c||c|c|c|c|}
\hline
$m_{1}$ & $m_{2}$ & $a_{1}$ & $a_{2}$ & deg & mult & $\{ u^{(1)}_{k} 
\}$ & $\{ u^{(2)}_{k} \}$\\   
\hline
0 & 0 & 2 & 0  & 10 & 1 & - & - \\
\hline
1 & 0 & 0 & 1  & 5 & 1 & 0.201347  & - \\
\hline
1 & 1 & 1 & 0  & 4 & 2 & 1.18368  &  1.35557\\
 &  &  &   &  &  & 0.18784  &  0.716566\\
\hline
2 & 2 & 0 & 0  & 1 & 2 & $0.844939  \pm 0.400816 i$  &  $1.07213  \pm 0.422759 i$\\
 &  &  &   &  &  & 0.211755\,, $1.48557 i$  &  $0.714804 i$\,, $2.1946 i$\\
\hline
\end{tabular}
\caption{\small $U_{q}(C_{2})$, $N=2$}\label{table:C2N2}
\end{table}

\begin{table}
\small
\centering
\begin{tabular}{|c|c||c|c||c|c|c|c|}
\hline
$m_{1}$ & $m_{2}$ & $a_{1}$ & $a_{2}$ & deg & mult & $\{ u^{(1)}_{k} 
\}$ & $\{ u^{(2)}_{k} \}$\\   
\hline
0 & 0 & 3 & 0  & 20 & 1 & - & - \\
\hline
1 & 0 & 1 & 1  & 16 & 2 & 0.115986  & - \\
 &  &  &   &  &  & 0.351133  & - \\
\hline
1 & 1 & 2 & 0  & 10 & 3 & 1.58467  & 1.70996 \\
 &  &  &   &  &  & 0.321003  & 0.760756 \\
 &  &  &   &  &  & 0.112316  & 0.701168 \\
\hline
2 & 1 & 0 & 1  & 5 & 3 & 0.382283\,, 0.963791  & 1.34441 \\
 &  &  &   &  &  & 0.118089\,, 1.05603  & 1.3902 \\
 &  &  &   &  &  & 0.113785\,, 0.333555  & 0.2923 \\
\hline
2 & 2 & 1 & 0  & 4 & 6 & 0.397606\,, 0.688759  & $0.9169  \pm 0.307663 i$ \\
 &  &  &   &  &  & $1.23957 \pm 0.466025 i$  & $1.39288 \pm 0.481069 i$ \\
 &  &  &   &  &  & 0.119249\,, $1.66217 i$  & $0.711593 i$\,, $2.20269 i$ \\
 &  &  &   &  &  & 0.116831\,, 0.865494  & $0.934114 \pm 0.250442 i$ \\
 &  &  &   &  &  & 0.385256\,, $1.71269 i$  & $0.61563 i$\,, $2.28518 i$ \\
 &  &  &   &  &  & 0.117124\,, 0.362471  & 0.34741\,, $2.68405 i$ \\
\hline
3 & 3 & 0 & 0  & 1 & 4 & 0.989238\,, $0.860023 \pm 0.700064 i$  & 
1.18442\,, $1.06721 \pm 0.745089 i$ \\
&  &  &   &  &  & 0.425069\,, 0.848958\,, $1.5453 i$  & 1.13679\,, 
$0.680288  i$\,, $2.20976 i$ \\
&  &  &   &  &  & 0.113851\,, 0.338372\,, $1.44183 i$  & $0.279454 
\pm 0.211351 i$\,, $2.39497 i$ \\
&  &  &   &  &  & 0.120953\,, 0.94247\,, $1.51754 i$  & 1.1893\,, 
$0.757077 i$\,, $2.14786 i$ \\
\hline
\end{tabular}
\caption{\small $U_{q}(C_{2})$, $N=3$}\label{table:C2N3}
\end{table}

\begin{table}
\small 
\centering
\begin{tabular}{|c|c|c||c|c|c||c|c|c|c|c|}
\hline
$m_{1}$ & $m_{2}$ & $m_{3}$ & $a_{1}$ & $a_{2}$  & $a_{3}$ & deg & mult & $\{ u^{(1)}_{k} 
\}$ & $\{ u^{(2)}_{k} \}$ & $\{ u^{(3)}_{k} \}$\\   
\hline
0 & 0 & 0 & 2 & 0 & 0  & 21 & 1 & - & - & - \\
\hline
1 & 0 & 0 & 0 & 1 & 0  & 14 & 1 & 0.201347  & - & - \\
\hline
1 & 1 & 1 & 1 & 0 & 0  & 6 & 2 & 0.190268  & 0.796966 & 1.04547 \\
 &  &  &  &  &   &  &  & 1.35599  & 1.55753 & 1.68531 \\
\hline
2 & 2 & 2 & 0 & 0 & 0  & 1 & 2 & $0.9507   \pm  0.448287 i$ & $1.21099  \pm 0.473011 i$  &  
$1.36739  \pm  0.485421 i$\\
 &  &  &  &  &   &  &  & 0.219256\,, $1.14053 i$  & $0.543566 i$\,, 
 $1.60164 i$ & $0.885636 i$\,, $2.04527 i$ \\
\hline
\end{tabular}
\caption{\small $U_{q}(C_{3})$, $N=2$}\label{table:C3N2}
\end{table}

\afterpage{%
    \clearpage
    \begin{landscape}
     
\small
\centering
\begin{tabular}{|c|c|c||c|c|c||c|c|c|c|c|}
\hline
$m_{1}$ & $m_{2}$ & $m_{3}$ & $a_{1}$ & $a_{2}$  & $a_{3}$ & deg & mult & $\{ u^{(1)}_{k} 
\}$ & $\{ u^{(2)}_{k} \}$ & $\{ u^{(3)}_{k} \}$\\   
\hline
0 & 0 & 0 & 3 & 0 & 0  & 56 & 1 & - & - & - \\
\hline
1 & 0 & 0 & 1 & 1 & 0  & 64 & 2 & 0.115986  & - & - \\
 &  &  &  &  &   &  &  & 0.351133  & - & - \\
\hline
1 & 1 & 1 & 2 & 0 & 0  & 21 & 3 & 0.113011  & 0.78232 & 1.03469 \\
 &  &  &  &  &   &  &  & 0.326182 & 0.839331 & 1.07712 \\
 &  &  &  &  &   &  &  & 1.86199 & 2.01428 & 2.10901 \\
\hline
2 & 1 & 0 & 0 & 0 & 1  & 14 & 1 & 0.115986\,, 0.351133  & 0.331791 & - \\
\hline
2 & 1 & 1 & 0 & 1 & 0  & 14 & 3 & 0.117608\,, 1.20956  & 1.58891 & 1.71382 \\
 &  &  &  &  &   &  &  & 0.11417\,, 0.336229 & 0.298556 & 0.751895 \\
 &  &  &  &  &   &  &  & 0.372422\,, 1.13284 & 1.54735 & 1.67608 \\
\hline
2 & 2 & 2 & 1 & 0 & 0  & 6 & 6 & 0.377177\,, 0.835326  & $1.03506 \pm 
0.320648i$ & $1.23102 \pm 0.37667 i$\\
 &  &  &  &  &   &  &  & $1.41135 \pm 0.538002i$  & $1.59277 \pm 
 0.555019 i$ & $1.70587 \pm 0.566549  i$ \\
 &  &  &  &  &   &  &  & 0.409958\,, $1.35937 i$ & $0.414206 i$\,, 
 $1.74096 i$ & $0.791369 i$\,, $2.1359 i$ \\
 &  &  &  &  &   &  &  & 0.116641\,, 0.980323 & $1.05068 \pm 0.260276 
 i$ & $1.24798 \pm 0.342251 i$ \\
 &  &  &  &  &   &  &  & 0.118386\,, 0.376545 & 0.367\,, $1.49213 i$ 
 & $0.65911 i$\,, $2.2416 i$ \\
 &  &  &  &  &   &  &  & 0.121155\,, $1.30785 i$ & $0.572787 i$\,, 
 $1.67608 i$ & $0.885614 i$\,, $2.04744 i$ \\
\hline
3 & 3 & 3 & 0 & 0 & 0  & 1 & 4 & 0.443222\,, 0.937877\,, $1.23794 i$ 
& 1.24501\,, $0.44955 i$\,, $1.66845 i$ & 1.41773\,, $0.832685 i$\,, 
$2.09128 i$\\
 &  &  &  &  &   &  &  & 1.11346\,, $0.953596 \pm 0.779036 i$ & 
 1.3403\,, $1.18988 \pm 0.830487 i$ & 1.47811\,, $1.32949 \pm 0.861223 
 i$ \\
 &  &  &  &  &   &  &  & ? & ? & ? \\
 &  &  &  &  &   &  &  & ? & ? & ? \\
\hline
\end{tabular}

	\captionof{table}{$U_{q}(C_{3})$, $N=3$}
    \label{table:C3N3}
    \end{landscape}
    \clearpage
}

\clearpage


\providecommand{\href}[2]{#2}\begingroup\raggedright\endgroup

\end{document}